\documentclass[10.8pt]{article}

\usepackage[round]{natbib}
\usepackage{amssymb, amsmath, graphicx, float, multirow, mathrsfs, amsfonts, enumitem}
\usepackage{subcaption}
\usepackage{bm}
\usepackage{comment}

\usepackage{color}

\newcommand\pr[1]{\text{Pr}\left(#1\right)}

\DeclareMathSymbol{\Y}{\mathbin}{AMSb}{"59}
\DeclareMathSymbol{\bS}{\mathbin}{AMSb}{"53}
\newcommand{\sci}{\bS}
\newcommand\sciimp{\sci^{\text{imp}}}  %  \textbf{y}^{\text{imp}}
\newcommand\Yobs{\textbf{Y}^{\text{obs}}}
\newcommand\yobs{\textbf{y}^{\text{obs}}}
\newcommand\tobs{{t}^{\text{obs}}}

\usepackage{amsthm}
\theoremstyle{definition}

\newtheorem{proposition}{Proposition}

\usepackage{authblk}

\title{A conditional randomization test to account for covariate imbalance in randomized experiments}
\author[*]{Jonathan Hennessy}
\author[*]{Tirthankar Dasgupta}
\author[*]{Luke Miratrix}
\author[**]{Cassandra Pattanayak}
\author[***]{Pradipta Sarkar}
\affil[*]{Department of Statistics, Harvard University}
\affil[**]{Quantitative Analysis Institute, Wellesley College}
\affil[***]{Principal Scientist, Procter \& Gamble International Operations}

\begin{document}

%\author{Hennessy, J., Dasgupta, T., Miratrix, L., Pattanayak, C. W., and Sarkar, P.}

\maketitle

\begin{abstract}
We consider the conditional randomization test as a way to account for covariate imbalance in randomized experiments.  The test accounts for covariate imbalance by comparing the observed test statistic to  the null distribution of the test statistic conditional on the observed covariate imbalance.  We prove that the conditional randomization test has the correct significance level and introduce original notation to describe covariate balance more formally. Through simulation, we verify that conditional randomization tests behave like more traditional forms of covariate adjustmet but have the added benefit of having the correct conditional significance level.  Finally, we apply the approach to a randomized product marketing experiment where covariate information was collected after randomization. 
\end{abstract}

% ----------------------------------------------------------------
\section{Introduction}
% ----------------------------------------------------------------

% Randomized experiments are the ``gold standard" for assessing causal effects.  Randomization
% , a ``reasoned basis for inference" \citep{Fisher1935},
% removes experimental bias and
In the context of randomized experiments, randomization allows for unbiased estimation of average causal effects and ensures that covariates will be balanced on average.
However, chance covariate imbalances do occur. To quote \cite{Senn1989},

\begin{quote}
  A frequent source of anxiety for clinical researchers is the process of randomization, and a commonly expressed worry, despite the care taken in randomization, is that the treatment groups will differ with respect to some important prognostic covariate whose influence it has proved impossible to control by design alone.
\end{quote}

\noindent For the imbalance to be an issue, the covariate needs to be prognostic (i.e. related to the outcome) but the covariate imbalance does not need to be statistically significant in order to affect the results \citep{Altman1985}.  Also, \citet{Senn1989} argued that in hypothesis testing, ``covariate imbalance is of as much concern in large studies as in small ones" because ``it is not the absolute imbalance which is important but the standardized imbalance and this is independent of sample size."

Restricted randomization and blocking are well-established strategies to ensure balance on key covariates.  More recently, \citet{Morgan2012} introduced rerandomization as a way to ensure balance on many covariates.  However, restricted randomization, blocking, and rerandomization are not always feasible.  In the product marketing example that motivated this work, the covariate information was not collected until after the units were assigned to treatment levels.  The experiment involved roughly 2000 experimental subjects and each subject randomly received by mail one of eleven versions of a particular product. Each subject used the product and returned a survey regarding the product's performance.  The outcome of interest was an ordinal variable with three levels, 1, 2, and 3, and the goal was to identify which product version the subjects preferred.  The survey also collected covariate information, such as income and ethnicity, and the experimenters were concerned about the influence of covariate imbalance on their conclusions.

While several methods exist to analyze ordinal data, including the proportional odds model, randomization tests are a natural choice because they require no assumptions about the distribution of the outcome.  Randomization tests are unique in statistics in that inference is completely derived from the physical act of randomization.  However, adjusting randomization tests for covariate imbalance is not straightforward.  To quote \citet{Rubin1980},

\begin{quote}
More complicated questions, such as those arising from the need to adjust for covariates brought to attention after the conduct of the experiment ... require statistical tools more flexible than FRTED (Fisher randomization tests for experimental data).
\end{quote}

There are two ways in which randomization tests can be used to adjust for covariate imbalance. One approach is to adjust the randomization test by modifying the test statistic, e.g., regressing the observed outcomes on the covariates and defining the test statistic in terms of the regression residuals. The second approach is to implement a conditional randomization test by conditioning on the covariate imbalance. In this article, we explore the second approach, i.e, conditioning as a way to adjust randomization tests for covariate imbalance.  The idea of conditioning is not new.
\citet{Rosenbaum1984} used these tests for inference on linear models with covariates.
% Conditional randomization tests have traditionally been used in the sequential design literature and only occasionally for covariate adjustment , Zheng2008}.
\citet{Zheng2008} proposed using the conditional randomization test to analyze multi-center clinical trials by conditioning on the number of treated subjects in each center.  They motivated the test primarily through simulations showing that the power of the conditional randomization test is greater than the power of the unconditional test.
%The emphasis on power, more precisely, unconditional power, is not surprising given that the usual rationale for covariate adjustment is increased precision and the results imply that conditioning on the observed covariate balance is similar to more traditional forms of covariance adjustment.
While \citet{Zheng2008} only considered the multi-center clinical trial, they were confident the idea could be applied more generally.
%, stating that ``Conditioning on the ancillary statistics in a randomization based analysis is a way of adjusting for the covariate effect,'' and that, ``This idea generalizes when there are an arbitrary number of covariates.'' \citet{Rosenbaum1984} also proposed a conditional randomization test  but in the context of an observational study.

In Section 2, we review the notation and basic mechanics of randomization tests.  In Section 3, we introduce conditional randomization tests and prove that the test has the correct significance level.  In Section 4, we apply the conditional randomization test to experiments with covariates. In Section 5, we evaluate the properties of the conditional randomization test via simulation and, in Section 6, we apply the test to the product marketing example.  In Section 7, we summarize our findings and lay out steps for future work.

% ----------------------------------------------------------------
\section{Randomization tests}
% ----------------------------------------------------------------

Randomization tests  \citep{Fisher1935}   for randomized experiments  have played a fundamental role in the theory and practice of statistics.  The early theory was developed by \citet{Pitman1938} and \citet{Kempthorne1952}.  In fact, \citet{Kempthorne1952} showed that many statistical procedures can be viewed as approximations of randomization tests.  To quote \citet{Bradley1968}, ``[a] corresponding parametric test is valid only to the extent that it results in the same statistical decision [as the randomization test].''

To introduce our notation and framework, we  briefly review the mechanics of randomization tests and prove that they are valid. This formulation will allow for more easily articulating the impact of conditioning later on. Consider a fixed sample of $N$ subjects or experimental units. Following Neyman (1923) (but see \cite{SplawaNeyman:1990ux}) and \cite{Rubin:1974wx}), let $Y_i(1)$ and $Y_i(0)$ be the potential outcomes for subject $i$ under treatment and control, respectively.
These are the outcomes we would see if we were to assign a unit to treatment or control, and are considered to be fixed, pre-treatment values.
Such a representation is adequate under the Stable Unit Treatment Value Assumption \citep{cox1958planning,rubin1980randomization}, called SUTVA, which states that there is only one version of the treatment and that there is no interference between subjects. We focus on finite sample inference, meaning we take the sample being experimented on as fixed.  Consequently, we can assemble all our potential outcomes into a ``Science Table'' that fully describes the sample. The Science Table is essentially a rectangular array denoted by $\sci$ in which each of the $N$ rows represents an experimental unit, the first two columns encode the two potential outcomes, and each of the remaining columns encode any covariates.

The individual or unit-level treatment effect for subject $i$ is then defined as a given comparison between $Y_i(1)$ and $Y_i(0)$. In particular, we focus on individual treatment effects of the form, $\tau_i = Y_i(1) - Y_i(0)$, though other comparisons are possible. Of course, we cannot observe both potential outcomes because we cannot simultaneously assign a unit to treatment and control. We instead observe $Y_i^{\text{obs}} = W_iY_i(1) + (1 - W_i)Y_i(0)$, where $W_i$ is the binary treatment assignment variable that takes the value 1 if unit $i$ is assigned to treatment and zero otherwise. We can record the entire assignment as a vector, $\bm{W} = (W_1, \ldots, W_N)$.
We also have the number of treated units $N_T = \sum_{i=1}^N W_i$   and the number of control units $N_C = N - N_T$.
In randomized marketing experiments that motivate our work, $N_C$ and $N_T$ are typically pre-fixed, although there are several examples of randomized experiments where it is not possible to pre-fix these quantities (e.g., in medical research).
The vector of observed outcomes $\Yobs$ can be written as  $\Yobs(\sci, \bm{W})$ to show its explicit dependence on $\sci$ and $\bm{W}$, and is random because of the randomness of $\bm{W}$.

%In practice, this means that identifying individual causal effects is only possible with strong assumptions. Therefore, researchers generally focus on effects either for some well-defined super-population or finite sample (see \cite{Rosenbaum:2012ul} for a discussion of finite sample causal inference).

We also have the assignment mechanism, $p(\bm{W})$, a distribution over all possible treatment assignments. We define $\mathscr{S}$, the set of \textit{acceptable treatment assignments}, as the set of all possible (allowed) assignment vectors $\bm{W} = (W_1, \ldots, W_N)$ for which $p(\bm{W}) > 0$. In most typical experiments, all treatment assignments in $\mathscr{S}$ are equally likely.
For instance, in the completely randomized design, $\text{p}(\bm{w}) = {N \choose N_T}^{-1}$ for any $\bm{w}$ such that $\sum w_i = N_T$.

Most randomization tests evaluate the Fisher sharp null hypothesis of no treatment effect:
\[ H_0: Y_i(1) = Y_i(0)  \mbox{ for } i = 1, \ldots, N . \]
%Importantly, under the sharp null $Y^{obs}$ is the same regardless of $W$.
To test this null, the experimenter first chooses an appropriate test statistic
\begin{equation}
t({\bm W}, \Yobs, \textbf{X}) \equiv t \left ( {\bm W}, \Yobs \left(\sci, \bm{W} \right), \textbf{X} \right), \label{eq:t}
\end{equation}
a function of the observed outcomes (and consequently of the Science table and the treatment assignment) and the covariates. Let $\bm{w}$ denote the observed assignment vector (realization of $\bm{W}$) and $\yobs$ denote the observed data (realization of $\Yobs$). The observed value
\begin{equation}
\tobs \equiv t({\bm w}, \yobs, \textbf{X}) \equiv t \left ( {\bm w}, \Yobs \left(\sci, \bm{w} \right), \textbf{X} \right) \label{eq:tobs}
\end{equation}
of the test statistic is then compared to its \emph{randomization distribution} under the sharp null.

To generate this randomization distribution, the missing potential outcome in each row of the Science Table is imputed with the observed value in that row, because under the sharp null the observed outcome and the missing outcome for any unit are equal. One therefore has a science table that is complete under the null hypothesis. This table can be used to obtain the null distribution of $t$ by calculating the value of $t$ from the outcomes that would be observed under each possible assignment vectors in $\mathscr{S}$. Finally, an \emph{extreme} (to be defined in advance by the experimenter) observed value of the test statistic with respect to its null distribution is taken as evidence against the sharp null, and eventually the sharp null is rejected if the observed value of the test statistic is larger than a pre-defined threshold. This can be formally described by the following four steps:

\begin{enumerate}

  \item Calculate observed test statistic, $\tobs \equiv t({\bm w}, \yobs, \textbf{X})$.

  \item Using $\bm{w}$, $\yobs$ and the sharp null hypothesis, fill-in the missing potential outcomes and denote the imputed potential outcomes table by $\sciimp$. Under the sharp null hypothesis of no treatment effect, $\sciimp = \sci$.

  \item Using $\sciimp$ and $\text{p}(\bm{W})$, find the reference distribution of the test statistic
  \begin{equation}
     t \left({\tilde{\bm W}}, \Yobs \left(\sciimp, \tilde{\bm{W}} \right), \textbf{X} \right) \equiv  t \left({\tilde{\bm W}}, \yobs, \textbf{X} \right), \label{eq:t_imp}
  \end{equation}
where ${\tilde{\bm W}}$ is a draw from $\text{p}(\bm{W})$. Note that (\ref{eq:t_imp}) holds because
$$ \Yobs \left(\sciimp, \tilde{\bm{W}} \right) \equiv \Yobs \left(\sci, \tilde{\bm{W}} \right) \equiv \yobs $$ by the equality of $\sciimp$ and $\sci$ under the sharp null hypothesis.

  \item Next we define the $p$-value, given an ordering of possible $t$ from less to more extreme.
  For example, using the absolute value of $t$ as the definition of extremeness, the $p$-value is
  \begin{equation}
    p = \text{Pr} \left( \left| t \left( \tilde{\bm W}, \yobs, \textbf{X} \right) \right|  \geq \left| \tobs \right| \right). \label{eq:pval}
  \end{equation}

  \item Reject the sharp null hypothesis if $p \leq \alpha$.

\end{enumerate}

\noindent Because $\mathscr{S}$ and $\text{p}(\bm{W})$ are used both to initially randomize the units to treatment and control and also to test the sharp null hypothesis, randomization tests follow the ``analyze as you randomize" principle due to \citet{Fisher1935}.

With the above description of the randomization test, it is straightforward to establish its validity, i.e., the fact that it has unconditional significance level $\alpha$. Let $U$ denote a random variable that has the same distribution as that of $\left| t \left( \tilde{\bm W}, \yobs, \textbf{X} \right) \right|$ and let $F_U(\cdot)$ denote the cumulative distribution function (CDF) of $U$. Then, successive application of (\ref{eq:pval}) and (\ref{eq:tobs}) yields
\begin{eqnarray*}
p = 1 - F_U \left( \left| \tobs \right| \right) = 1 - F_U \left( \left | t \left ( {\bm w}, \Yobs \left(\sci, \bm{w} \right), \textbf{X} \right) \right|, \right)
\end{eqnarray*}
The distribution of $p$ over all possible observed randomizations is the same as the distribution of $$1 - F_U \left( \left| t \left( {\bm W}, \Yobs \left(\sci, \bm{W} \right) ,\textbf{X} \right) \right| \right),$$ which, under the sharp null hypothesis has the same distribution as that of $1 - F_U(U)$ by the equivalence of $\left| t \left( {\bm W}, \Yobs \left(\sci, \bm{W} \right) ,\textbf{X} \right) \right|$ and $ \left| t \left( {\bm W}, \yobs, {\bm X} \right) \right|$. Consequently, by the probability integral transformation, $p$ has a uniform $[0,1]$ distribution under the sharp null, and it follows that
$$ Pr \left(p \le \alpha | H_0 \right) \le \alpha, $$ proving that the randomization test has unconditional significance level $\alpha$.

% ----------------------------------------------------------------
\section{Conditional randomization tests}
% ----------------------------------------------------------------

We begin the discussion of conditional randomization tests by reviewing some history and arguing that they are appropriate to account for covariate imbalance observed after the experiment is conducted.
While \citet{Cox1982} introduced the conditional randomization test, the idea of conditional inference can be traced back to Fisher and his notion of relevant subsets \citep{Fisher1956}.  Conceptually, testing the null of $\theta = \theta_0$ for some parameter is done by comparing the observed data to hypothetical observations that might have been observed given $\theta_0$. To do this, we need to select the sets of  hypothetical observations that should be used as a point of comparison.
Fisher believed this set should not necessarily include all hypothetical observations and should be chosen carefully.  He called this set the relevant subset of hypothetical observations.  To quote \citet{Cox1958}, relevant subsets

\begin{quote}
should be taken to consist, so far as is possible, of observations similar to the observed set in all respects which do not give a basis for discrimination between possible values of the unknown parameter of interest.
\end{quote}

\noindent The idea of ``observations similar to the observed set" is admittedly vague, and it is not immediately obvious why a subset of the hypothetical observations should lead to better inferences.  The idea and its implications have been extensively studied and debated in the statistics literature.  See, for example, \citet{Cox1958}, \citet{Kalbfleisch1975}, and \citet{Helland1995}. However, certain principles have become well established and we focus on those.

Relevant subsets are closely related to ancillary statistics.  By definition, the distribution of ancillary statistics do not depend on the unknown parameter of interest.  Also, observations with the same value of the ancillary statistic share some similarity to each other.  Because ancillary statistics do not depend on the parameter of interest, different observations with the same value of the ancillary statistic should not favor one parameter value over another.
Thus, such observations form a relevant subset. The temperature testing example by \citet{Cox1958} is perhaps the best known example of this idea.
\citet{Birnbaum1962} formalized this notion as the \textit{conditionality principle}.  The conditionality principle applies when running an experiment $E$ by first randomly selecting one of several component experiments  $E_1, \ldots, E_m$ and, second, running the selected experiment.  The conditionality principle says that the \textit{evidential meaning} of the experiment is the same as the meaning of the randomly selected component experiment.
As \citet{Kalbfleisch1975} put it, which experiment was selected is an \textit{experimentally ancillary statistic}. More colloquially, ``any experiment not performed is irrelevant" \citep{Helland1995}.
%Cox's example fits directly into this context because the two thermometers represent two component experiments.  \citet{Kalbfleisch1975} called the selected experiment an \textit{experimentally ancillary statistic}.
Overall, this suggests that we compare what we have to the distribution of what we would have had under the null, given that any ancillary (unrelated) pieces of information (such as realized number of units treated) matches.

% ----------------------------------------------------------------
\subsection{Conditional randomization test mechanics}
% ----------------------------------------------------------------

Our development of the conditional randomization test parallels \citet{Kiefer1977}'s development of the conditional confidence methodology, especially the notion of partitions.  Let $\mathscr{S}_1, \ldots, \mathscr{S}_m$ partition the set of acceptable treatment assignments, $\mathscr{S}$, such that $\mathscr{S}_i \cap \mathscr{S}_j = \emptyset$ for all $i \neq j$ and $\cup_{i=1}^{m} \mathscr{S}_i = \mathscr{S}$.
Then for any observed and allowed random assignment $w$, define $\mathscr{S}({\bm w})$ as the (unique) partition containing ${\bm w}$.
 We shortly discuss different ways in which $\mathscr{S}_1, \ldots, \mathscr{S}_m$  are constructed, but for now, assume that the partitions as given.

\begin{comment}
Let $\pi_i = \text{Pr}(\bm{W} \in \mathscr{S}_i)$ be the probability of selecting a treatment assignment from the $i$th partition.  Selecting the treatment assignment according to $\text{p}(\bm{W})$ is equivalent to first selecting one of the partitions using the probabilities $(\pi_1, \ldots, \pi_m)$ and then selecting the treatment assignment from the partition according to $\text{p}(\bm{W} \, | \, \bm{W} \in \mathscr{S}_i)$, where

\begin{equation}
  \text{Pr}(\bm{W} = \omega \, | \, \bm{w} \in \mathscr{S}_{i}) = \frac{\text{Pr}(\bm{W} = \omega)}{ \sum_{\omega' \in \mathscr{S}_{i}} \text{Pr}(\bm{W} = \omega') } \textbf{1}_{\omega \in \mathscr{S}_{i}}.
\end{equation}
\end{comment}

Thus, we can frame this experiment as a mixture of component experiments, where each partition corresponds to a component experiment.  Following the conditionality principle, we should then only consider the selected partition of treatment assignments when carrying out the test.

In a conditional randomization test, we define the ``reference set'' $\mathscr{S}_{\text{ref}}$
%= \mathscr{S}({\bm w})$
as the partition that contains the observed treatment assignment. Then we use $\mathscr{S}_{\text{ref}}$ to generate draws from the randomization distribution. We emphasize this by writing $\mathscr{S}_{\text{ref}} = \mathscr{S}_{\text{ref}}(\bm{w})$. Consequently, conditional randomization tests do not entirely follow the ``analyze as you randomize" principle. It is worthwhile to note here that in the unconditional randomization test, the reference set $\mathscr{S}_{\text{ref}}$ is the same as the set $\mathscr{S}$ of all acceptable treatment assignments.

As we did for randomization tests, we lay out the steps of the conditional randomization test.
Given an observed treatment assignment, $\bm{W} = \bm{w}$, from $\mathscr{S}$ and, observed $\Yobs = \yobs$, take the following steps:
\begin{enumerate}

  \item Calculate observed test statistic, $\tobs \equiv t({\bm w}, \yobs, \textbf{X}) \equiv t \left ( {\bm w}, \Yobs \left(\sci, \bm{w} \right), \textbf{X} \right)$.

  \item Using $\bm{w}$, $\yobs$, and the sharp null hypothesis, impute the potential outcomes table $\sciimp$, which equals $\sci$ under the sharp null.

  \item Using $\sciimp$ and $\text{p}(\bm{W} \, | \, \bm{W} \in \mathscr{S}_{\text{ref}}(\bm{w}))$, find the conditional reference distribution of the test statistic
  $ t \left( \tilde{\bm W}, \Yobs\left( \sciimp, \tilde{\bm W} \right), {\textbf X} \right) \equiv t \left({\tilde{\bm W}}, \yobs, \textbf{X} \right)$, given that $\tilde{\bm W} \in \mathscr{S}_{\text{ref}}(\bm{w})$.
%  \begin{equation}
%    \text{p}(t(\bm{w}, \yobs(\bm{w}, \sciimp), \textbf{X}) \, | \, \bm{w} \in \mathscr{S}_{\text{ref}}(\bm{w})).
%  \end{equation}

  where ${\tilde{\bm W}}$ is a draw from $\text{p}(\bm{W})$.

  \item Next we define the $p$-value as:
  \begin{equation}
    p = \text{Pr} \left( \left| t \left( \tilde{\bm W}, \yobs, \textbf{X} \right) \right|  \geq \left| \tobs \right| \Big| \tilde{\bm W} \in \mathscr{S}_{\text{ref}}(\bm{w}) \right). \label{eq:pvalcond}
  \end{equation}

  \item Reject the sharp null hypothesis if $p \leq \alpha$.

  \end{enumerate}

%In the following, we discuss how to create partitions to accommodate and account for covariate adjustments.

%In any experiment, the experimenter will likely be faced with several different ways to form the partitions.  While no general theory exists, according to \citet{Berger1988}, the general idea is to ``find subsets ... which when conditioned upon, change the experimental measure."  In our setting, this refers to the conditional properties of the test differing from the unconditional properties of the test.  If the conditional and unconditional properties differ, then the subset should be conditioned on.

% ----------------------------------------------------------------
\subsection{The validity of the conditional randomization test}
% ----------------------------------------------------------------
The conditional randomization test is valid if the test unconditionally rejects the sharp null with probability $\leq \alpha$.
We show this now through an argument similar to the one used for establishing the validity of the unconditional randomization test. Define a sequence of $p$-values $p_1, \ldots, p_m$, where
\begin{equation}
    p_i = \text{Pr} \left( \left| t \left( \tilde{\bm W}, \yobs, \textbf{X} \right) \right|  \geq \left| \tobs \right| \Big| \tilde{\bm W} \in \mathscr{S}_i) \right), \label{eq:pval_i}
\end{equation}
and define the rejection rule as $p_i \le \alpha$ if our observed randomization $w$ is in $\mathscr{S}_i$. Then, the probability of rejecting the sharp null hypothesis when it is true is:
\begin{eqnarray*}
&& \sum_{i=1}^m Pr_{H_0} \left( p_i \le \alpha \right | {\bm W} \in \mathscr{S}_i ) Pr ({\bm W} \in \mathscr{S}_i ) \\
&\le& \sum_{i=1}^m \alpha Pr ({\bm W} \in \mathscr{S}_i ), \;\ \mbox{by the validity of the unconditional randomization test}, \\
&=& \alpha.
\end{eqnarray*}

Thus, the conditional randomization test has unconditional significance level $\alpha$. There are some restrictions on the partitions, $\mathscr{S}_1, \ldots, \mathscr{S}_m$.  For a given partition, $\mathscr{S}_i$, in order for the $p$-value to ever be $\leq \alpha$, the number of elements in $\mathscr{S}_i$ must be $\geq \alpha^{-1}$.  Otherwise, even the most extreme value of the test statistic would not lead to the sharp null being rejected.

Additionally, in order for the test to have significance level $\alpha$, the partitions must be specified before the experimenter has access to the observed outcomes.  Otherwise, the experimenter could consciously or subconsciously manipulate the inference by changing the reference distribution.  This follows Rubin's principle of separating design from analysis; see, for example, \citet{Rubin2007}.

% -----------------------------------------------------------------------------------------------
\section{Implementation of conditional randomization tests: partitioning of treatment assignments and test statistics}
% -----------------------------------------------------------------------------------------------

Having described the conditional randomization test and its mechanics, we now need to address the following issues:
\begin{enumerate}
\item How to partition the set of acceptable treatment assignments. Since our research was motivated by the need to adjust for covariate imbalance across treatment groups observed after conducting the experiment, a natural strategy is to use a measure of covariate balance across treatment groups as a partitioning variable (or variables).  We discuss how to do this.
\item How to select a test statistic to use for the conditional randomization test.
 For example, should the test statistic be adjusted for covariate imbalance by regressing the observed outcome on the covariates and re-defining it in terms of regression residuals, as done by \cite{Rosenbaum2002}?
\end{enumerate}

\subsection{Partitioning of treatment assignments using a covariate balance function}

The overall logic behind using covariate balance to partition treatment assignments is simple: in a balanced randomization, even small deviations of the test statistic will tend to be relatively rare and we should reject accordingly if they are observed.  In an imbalanced randomization, however, it is easier to have extreme values of the statistic, so we should not reject in such circumstances.  Thus the location and spread of the reference distribution should
%be more narrow or broad to
reflect this. We first illustrate this aspect with an example. Consider an experiment with $N = 100$ units assigned according to a completely randomized design where $N_T = N_C = 50$. Let the sharp null hypothesis of no treatment effect be true and the test statistic be $t = \bar{Y}^{\text{obs}}_T - \bar{Y}^{\text{obs}}_C$, where $\bar{Y}^{\text{obs}}_T$ and $\bar{Y}^{\text{obs}}_C$ respectively denote the average observed outcomes of units exposed to treatment and control.
We observe some continuous outcome such as health.
We also observe the covariate of the units' sex: there are 50 males and 50 females.
For the sake of the example, assume that males tend to have higher potential outcomes than females.

The experimenter assigns units to treatment and control but ends up with an unbalanced treatment assignment with $N_{T1} = 35$ men in the treatment group and $N_{C1} = 15$ men in the control group.
This covariate imbalance creates complications: males and females have different potential outcome distributions and so even under the null we would expect a positive difference in the groups.
At this point, the experimenter knows that the probability of rejecting the sharp null is much higher than $0.05$.

\begin{figure}[htbp]
  \begin{center}
    \includegraphics[width=3.5in]{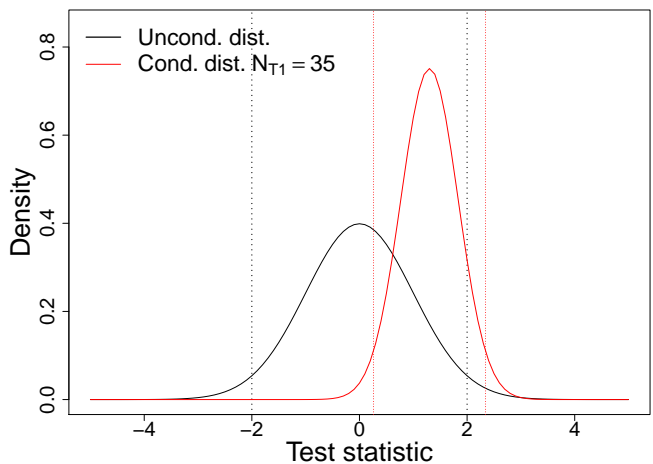}
  \end{center}
  \caption{\small \textbf{Unconditional and conditional distributions of test statistic:}  The unconditional distribution is the black solid line and the black vertical dotted lines mark the unconditional rejection region.  The conditional distribution when $N_{T1} = 35$ is the solid red and the red vertical lines mark the conditional rejection region.  The conditional probability of rejecting the test using the unconditional rejection region is $0.21$.}
  \label{simple_comp_cov}
\end{figure}

This is illustrated on Figure~\ref{simple_comp_cov}.
The unconditional distribution of the test statistic is the solid black line and the black dotted lines at $-2$ and $2$ mark the rejection region for the unconditional test.  The unconditional probability the experimenter observes a test statistic in the rejection region is $0.05$.
The distribution of the test statistic conditioned on $N_{T1}$ however, is the red line; the probability of being less than $-2$ or greater than $2$ is $0.2$.
The red dotted lines mark the conditional rejection region based on the conditional distribution.
Now, given $N_{T1} = 35$, the experimenter faces a choice: use the unconditional test, knowing the randomization went poorly, or use the conditional test and have conditionally valid results.
We believe the latter choice is correct; it is essentially adjusting the test based on the distribution of the covariates.
This is philosophically similar to the practice of using the covariates to construct an adjusted test statistic \citep{Rosenbaum1984}.

We construct a conditioning partition by grouping potential assignment vectors using similarity on ``balance.''
To do this, we first need a measure of balance, which we formalize now.
Let the covariate balance function $B(\bm{w}, \textbf{X})$ be a function of $\bm{w}$ and $\textbf{X}$.  The covariate balance function reports a relevant summary of the covariate distribution for each level of the treatment.  For instance, if the mean and variance are appropriate summaries of the covariate distribution, the covariate balance function should report the mean and variance of each covariate for each  treatment level.

We can use the covariate balance function to partition the set of treatment assignments.  Let $\mathscr{B}$ be the set of all possible values of covariate balance function.  For each $b \in \mathscr{B}$, let $\mathscr{S}_b = \lbrace \omega \, : \, B(\omega, \textbf{X}) = b \rbrace$ be the set of treatment assignments with the same value of the covariate balance function, where $\cup_{b \in \mathscr{B}} \mathscr{S}_b = \mathscr{S}$.  We carry out the conditional randomization test using these partitions.

For categorical covariates, we can define the covariate balance function in terms of the cells of a contingency table where the rows are the levels of the covariate and the columns are the treatment levels.  We start with the case of a single categorical covariate with $J$ levels and a treatment with $K$ levels, visualized in Table \ref{cont_table}.
%The row sums are always fixed and, for a completely randomized design, the column sums are fixed.
 A natural covariate balance function is the contingency table itself (i.e. the matrix of internal cells, $\lbrack N_{j,k} \rbrack$). Thus, $B(\bm{w}, \textbf{X}) = \lbrack N_{j,k} \rbrack$ and if $B(\bm{w}, \textbf{X}) = b$, then $\mathscr{S}_b$ is made up of those treatment assignments that produce contingency table $b$.

\begin{table}[ht]
\caption{ \textbf{Single categorical covariate:} For the case of one categorical covariate, the contingency table summarizes the distribution of the covariate in each level of the treatment.  For a completely randomized design, a natural covariate balance function is the matrix of internal cells.}
\begin{center}
  \begin{tabular}{cr|cccc|c}
& & \multicolumn{4}{c|}{$W$} &  \\
& & 1 & 2 & $\cdots$ & $K$ &  \\
  \hline
\multirow{4}{*}{$X$} & 1 & $N_{1,1}$ & $N_{1,2}$ & $\cdots$ & $N_{1,K}$ & $N_{1, \cdot}$ \\
 & 2 & $N_{2,1}$ & $N_{2,2}$ & $\cdots$ & $N_{2,K}$ & $N_{2, \cdot}$\\
 & $\vdots$ & $\vdots$ & $\vdots$ & $\ddots$ & $\vdots$ & $\vdots$ \\
 & $J$ & $N_{J,1}$ & $N_{J,2}$ & $\cdots$ & $N_{J,K}$ & $N_{J, \cdot}$ \\
 \hline
 & & $N_{\cdot, 1}$ & $N_{\cdot, 2}$ & $\cdots$ & $N_{\cdot, K}$ & $N_{\cdot, \cdot}$
\end{tabular}
\end{center}
\label{cont_table}
\end{table}

We can also use the contingency table to discuss the covariate balance function when there are multiple categorical covariates.  The combinations of the categorical covariates (i.e. the Cartesian product) can be treated as the levels of a single categorical covariate.  As an example, consider the case of two binary categorical covariates, $X_1$ and $X_2$, and a binary treatment.  The contingency table considering all combinations of the covariates is shown in Table \ref{mult_cov}.

\begin{table}[ht]
\caption{\textbf{Multiple categorical covariates:} For the case of two categorical covariates, the combinations of the two categorical covariates can be treated as the levels of a single categorical covariate.}
\begin{center}
  \begin{tabular}{r|cc|c}
 & \multicolumn{2}{c|}{$W$} &  \\
 & 0 & 1 &  \\
  \hline
 $X_1 = 0, \, X_2 = 0$ & $N_{00, 0}$ & $N_{00, 1}$ & $N_{00, \cdot}$ \\
 $X_1 = 0, \, X_2 = 1$ & $N_{01, 0}$ & $N_{01, 1}$ & $N_{01, \cdot}$\\
 $X_1 = 1, \, X_2 = 0$ & $N_{10, 0}$ & $N_{10, 1}$ & $N_{10, \cdot}$\\
 $X_1 = 1, \, X_2 = 1$ & $N_{11, 0}$ & $N_{11, 1}$ & $N_{11, \cdot}$\\
 \hline
  & $N_{\cdot \cdot, 1}$ & $N_{\cdot \cdot, 2}$ & $N_{\cdot\cdot, \cdot}$
\end{tabular}
\end{center}
\label{mult_cov}
\end{table}

In this case, we could let the covariate balance function be the contingency table.  However, such a covariate balance function implies that the interaction between $X_1$ and $X_2$ is as important as $X_1$ and $X_2$ individually.  While plausible in some contexts, the interaction is generally less prognostic.  The number of units with $X_1 = 1$ assigned to treatment and the number of units with $X_2 = 1$ assigned to treatment are typically of greater interest.  We therefore might instead use a covariate balance function of

\begin{equation}
  B(\bm{w}, \textbf{X}) = (N_{10, 1} + N_{11, 1}, N_{01, 1} + N_{11, 1}).
\end{equation}

\noindent where $N_{10, 1} + N_{11, 1}$ are the number of units assigned to treatment with $X_1 = 1$  and $N_{01, 1} + N_{11, 1}$ are the number of units assigned to treatment with $X_2 = 1$.  If $B(\bm{w}, \textbf{X}) = b$, $\mathscr{S}_b$ consists of treatment assignments that produce the observed contingency table as well as treatment assignments that produce different contingency tables consistent with the marginal balance function $B(\bm{w}, \textbf{X}) = b$.

The covariate balance function could also make use of a cluster analysis or other methods of dimension reduction.  In a cluster analysis, observations are assigned to clusters such that the observations within each cluster are more similar to each other than to those observations in other clusters.  Popular clustering methods include $k$-means for continuous variables and $k$-modes for categorical variables \citep{Huang1997a}.  Clustering methods also exist for data sets with both continuous and categorical variables \citep{Wilson1997, McCane2008}.  Once the clusters have been formed, the covariates can be replaced with a single categorical covariate indicating cluster membership.  The covariate balance function would then be the number of treated units within each cluster.
We explain this approach further using the example of our marketing experiment in Section 6.

% One of the hazards of continuous covariates is that, to quote \citet{Rosenbaum1984}, ``the number of treatment assignments [in a partition] may be too small to be of practical use."  As mentioned earlier, if $|\mathscr{S}_b | < \alpha^{-1}$, the size of the conditional test will be greater than $\alpha$.

Many categorical or truly continuous variables will give partitions containing very few, or even only one, possible treatment assignment.
Recall from earlier that, if we want any power, we require $|\mathscr{S}_b | > \alpha^{-1}$ to allow for the size of the conditional test to be bounded by $\alpha$.
For continuous covariates one possible remedy is to coarsen (i.e. round) the continuous covariates such that there are enough treatment assignments with the same covariate balance.  For example, one might create income buckets, such at \$20,000-\$40,000, etc.
Such an approach destroys some information but hopefully not too much if carried out with the help of a subject matter expert.  This is reminiscent of Coarsened Exact Matching \citep{Iacus2012}, in which all covariates are discretized and balance is described by the number of units in each combination of the categorical covariates for each treatment level.  Because the covariates in our motivating example are all categorical, we focus on the categorical covariate case and leave the continuous case for future work.

% ----------------------------------------------------------------
\subsection{Choice of test statistic}
% ----------------------------------------------------------------

 A common test statistic in two-level randomized experiments (e.g., treatment-control studies) is the \emph{simple difference} of the observed outcomes in the treatment and control groups, i.e.
 \begin{equation}
 t_{\text{sd}} = \bar{Y}_T^{\text{obs}} - \bar{Y}_C^{\text{obs}}, \label{eq:t_sd}
 \end{equation}
 where $\bar{Y}_T^{\text{obs}}$ and $\bar{Y}_C^{\text{obs}}$ respectively denote the average observed responses in the treatment and control group respectively. A standardized version of $t_{\text{sd}}$ can also be used. However, keeping in mind the alternative strategy of adjusting randomization tests for covariate imbalance by modifying the test statistic, one may be tempted to use such adjusted test statistic for conditional randomization tests as well.

 A popular method of adjusting randomization tests for covariate imbalance is to first regress the observed potential outcomes on the covariates.  The residuals from the regression are treated as the ``adjusted outcomes" and the randomization test is carried out by calculating the test statistic using the adjusted outcomes in place of the observed potential outcomes.  For instance, if $Y_i^{\text{obs}}$ is continuous we can let the residuals be

\begin{equation}
  e_i^{\text{obs}} = Y_i^{\text{obs}} - f(X_i)
\end{equation}

\noindent where $f(\cdot)$ is a flexible, potentially non-parametric, function that does not depend on $Y$ under the null.  The test statistic can be, for instance, the difference between the mean of the residuals in the treatment and control group,

\begin{equation}
  t(\bm{W}, \Yobs, \textbf{X}) = \bar{e}_T^{\text{obs}} - \bar{e}_C^{\text{obs}}. \label{eq:t_res}
\end{equation}

This approach is described in both \citet{Raz1990} and \citet{Rosenbaum2002}.  \citet{Tukey1993} also described a similar procedure but recommended first creating ``compound covariates," typically linear combinations of existing covariates, and using the compound covariates in the regression, which is similar in spirit to principal component regression.  If the outcome is discrete, \citet{Gail1988} proposed using components of the score function derived from a generalized linear model as the adjusted outcome.

When the covariate is categorical, this adjustment is often called post-stratification and we refer to the levels of the covariate as strata.  \citet{Pattanayak2011} and \citet{Miratrix2013} studied post-stratification from the Neymanian perspective and derived the unconditional and conditional distributions of two estimators. The post-stratified estimate of treatment effect (which can be used as a test statistic) is defined as
\begin{equation}
  t_{\text{ps}} = \sum_{j = 1}^{J} \frac{N_j}{N} t_{\text{sd}, j}, \label{eq:t_ps}
\end{equation}
where $t_{\text{sd},j}$ denotes the standard test statistic given by (\ref{eq:t_sd}) for the $j$th stratum for $j=1, \ldots, J$.
%This post-stratified test statistic is the same as the simple difference in means of the residuals obtained by regressing the observed outcomes on the covariate.
%\noindent While $\hat{\tau}_{sd}$ ignores the covariate information, $\hat{\tau}_{ps}$ utilizes it; it is an example of adjusting for covariate imbalance through choice of test statistic.

We now state a result which shows that the conditional randomization tests using $t_{\text{sd}}$ and $t_{\text{ps}}$ are equivalent, if there is one categorical covariate.

\begin{proposition} \label{prop1}
Let $X$ denote a categorical covariate with $J$ levels, observed after a two-armed randomized experiment is conducted with $N$ units. Let $N_j$ denote the observed number of units that belong to stratum $j$, and let $N_{Tj}$ and $N_{Cj}$ denote the number of units assigned to treatment and control respectively, in stratum $j$, such that $N_{Tj} + N_{Cj} = N_j$, and $\sum_{j=1}^J N_j = N$. Then the conditional randomization test using the standard test statistic $t_{\text{sd}}$ defined by (\ref{eq:t_sd}) and the balance function $(N_{TI}, \ldots, N_{TJ})$ is equivalent to the unconditional randomization test using the composite test statistic $t_{\text{ps}}$ defined by (\ref{eq:t_ps}).
\end{proposition}

Proposition \ref{prop1} can be proved by adapting a proof from \citet{Rosenbaum1984}, and arguing that $t_{\text{ps}}$ is a monotonic function of $t_{\text{sd}}$. Please refer to Appendix A for details. It is worthwhile to note that the fact that $t_{sd}$ and $t_{ps}$ leads to the same conditional randomization test procedure can be intuitively understood from the fact that $t_{ps}$ itself can be viewed as a ``conditional estimator.'' Also, the equivalence of the two procedures does not necessarily mean that they are equally advantageous and disadvantageous under all situations. Using $t_{ps}$ has some advantages. For example, \citet{Ding2014} showed that asymptotic Neymanian inference sometimes gives more powerful tests, and thus using $t_{ps}$ and its Neymanian variance to test the null hypothesis may be a better choice	in	terms of power.	On the other hand, $t_{ps}$ may be disadvantageous to use when the number of categories of the discrete covariate is large because $t_{ps}$ has bad repeated sampling properties with finite samples. However, the conditional randomization test does not suffer from this problem, because one does not have to choose the balance function $(N_{T1}, \ldots, N_{TJ})$. Further, conditional randomization test statistics can be general, e.g., rank-based statistics, but it is not straighforward to obtain analogues of such test statistics for $t_{ps}$.

We conclude this Section with the remark that using a conditional randomization test or using a randomization test with an adjusted statistic are both more robust strategies than ANCOVA, which involves regressing $\yobs$ on $\bm{w}$ and $\textbf{X}$ and testing the treatment effect by carrying out a $t$ or $F$ test for the inclusion of $\bm{w}$, because randomization-based methods do not assume that the model is correctly specified. The nominal size for the randomization test using the residuals is maintained even when relevant covariates are not included in the regression and the assumed distribution for the outcome is incorrect. \citet{Stephens2013} carried out an extensive simulation study to compare such randomization tests to model-based regression approaches, including \citet{Zhang2008}'s semi-parametric estimator.  They found that the model based approaches often inflate the probability of Type I error, whereas permutation methods do not.

% ----------------------------------------------------------------
\section{Simulation Study}
% ----------------------------------------------------------------

We next illustrate via simulation the unconditional and conditional properties of the conditional randomization test as compared to two unconditional randomization tests.  For this simulation, the relevant unconditional properties of the tests are the average rejection rates over repeated runs of the experiment.  The conditional properties of the test are the average rejection rates
% under repeated runs of the experiment
where the covariate balance is held fixed.  For a given experiment, the conditional rejection rates are arguably more relevant than the unconditional rejection rates.  While the unconditional rejection rates measure the performance of the test over all treatment assignments, the conditional rejection rates measure the performance of the test for treatment assignments like the observed one.

%To again quote \citet{Cox1958},
%
%\begin{quote}
%Our calculation of power, etc. should be made conditionally within the distribution known to have been sampled.
%\end{quote}

%We utilize many of their results in evaluating the properties of unconditional and conditional randomization tests.

%Let $\textbf{X} = (X_1, \ldots, X_N)$ be the vector of strata indicators and let $J$ be the number of strata.  The units in strata $j$ are given by $\nu_j = \lbrace i : X_i = j \rbrace$.
We examine a completely randomized design with two treatment levels and a categorical covariate $B_i \in \left\{1, \ldots, J\right\}$.
Define dummy variables $X_{ij}$ with $X_{ij} = 1$ if unit $i$ is in stratum $j$.
 $N_T$ units are assigned to treatment and $N_C = N - N_T$ units are assigned to control.
 Let the covariate balance function be the number of treated units in the strata,
\begin{align}
%  B(\bm{w}, \textbf{X}) &= \Big(\sum_{i \in \nu_1} W_i, \ldots, \sum_{i \in \nu_J} W_i\Big) \nonumber \\
  B(\bm{w}, \textbf{X}) &= (N_{T1}, \ldots, N_{TJ}),
\end{align}
with $N_{Tj}$ the number of treated units and $N_j$ the number of units in the $j$th stratum.
Then $N_T = \sum_{j=1}^J N_{Tj}$ and $N_C = \sum_{j=1}^J N_{Cj}$.

% Let $\bar{Y}_j^{\text{obs}}$ be the observed mean outcome in the $j$th stratum,
%
%\begin{equation}
%  \bar{Y}_j^{\text{obs}} = \frac{1}{N_j} \sum_{B_i = j} Y_i^{\text{obs}}
%\end{equation}
%
%\noindent and $\bar{Y}^{\text{obs}} = \frac{1}{N}\sum_{j=1}^J N_j  \bar{Y}_j^{\text{obs}}$.

We compare the conditional and unconditional randomization tests over several simulation settings and with both $\widehat{\tau}_{sd}$, the simple difference statistic, and $\widehat{\tau}_{ps}$, the post-stratified test statistic.  Since the conditional randomization tests with $\widehat{\tau}_{sd}$ and $\widehat{\tau}_{ps}$ are equivalent, we only report results for the conditional randomization test using $\widehat{\tau}_{sd}$.   We let $N = 100$, $N_T = 50$, and $N_C = N - N_T = 50$.  We also let the number of strata be $J = 2$ and $N_1 = N_2 = 50$.  See Table~\ref{sim_ps}.
 Because there are only two strata and two treatment levels, the covariate balance function is completely determined by $N_{T1}$, the number of treated units in the first stratum.

\begin{table}[h]
  \caption{\textbf{Simulation design:} We use a completely randomized design where $N = 100$ and $N_T = 50$.}
  \begin{center}
  \begin{tabular}{cl|cc|c}
    	& & \multicolumn{2}{c|}{$W$} & \\
			& & $1$ & $0$ & \\
			\hline
			\multirow{2}{*}{$X$} & 1 & $N_{T1}$ & $N_{C1}$ & $N_1 = 50$ \\
			 & 2 & $N_{T2}$ & $N_{C2}$ & $N_2 = 50$ \\
			\hline
			& & $N_T = 50$ & $N_C = 50$ & $N = 100$ \\
		\end{tabular}
    \end{center}
    \label{sim_ps}
\end{table}

We generate the ``science'', the complete potential outcomes table, by varying two parameters, $\tau$ and $\lambda$.  Here, $\tau$ is the additive unit-level treatment effect and $\lambda$ measures the association between $X$ and $Y(0)$ (i.e. the prognostic ability of $X$).

\begin{align}
  \tau &= Y_i(1) - Y_i(0) \nonumber \\
  \lambda &= E(Y(0) \, | \, X = 2) - E(Y(0) \, | \, X = 1)
\end{align}

\noindent We let $\tau$ take on one of $11$ values, $\tau \in \{0, 0.1, 0.2, \ldots, 1 \}$ and $\lambda$ take on one of three values, $\lambda \in \{0, 1.5, 3 \}$.  We generate the complete potential outcomes by first drawing $Y_i(0) \, | \, X_i$ and then filling in $Y_i(1)$ as follows.

\begin{align}
  Y_i(0) \, | \, X_i &\sim N(\lambda X_i, 1) \nonumber \\
  Y_i(1) &= Y_i(0) + \tau
\end{align}

\noindent After generating the potential outcomes, we randomly assign units to treatment and control and record whether each of the three tests (two unconditional tests and one conditional test) rejects the sharp null, $H_0: Y_i(1) = Y_i(0)$ for $i = 1, \ldots, N$, at the 0.05 significance level.  We repeat this 1000 times and record the average rejection rate for each test.

We randomly assign the units in one of two ways.  We either assign them using the completely randomized assignment mechanism or we assign them holding $N_{T1}$ fixed at either 25, 30, 35, or 40.  Assigning the units using the completely randomized assignment mechanism allows us to evaluate the unconditional properties of the test and holding $N_{T1}$ fixed allows us to assess the conditional properties of the test (i.e. how the test performs for particular values of $N_{T1}$). Since we are implicitly interested in situations where the covariate is prognostic, when evaluating the conditional properties, we let $\lambda = 3$.

% ----------------------------------------------------------------
\subsection{Unconditional properties}
% ----------------------------------------------------------------

\begin{figure}[H]
\begin{center}
\begin{minipage}[b]{0.45\linewidth}
  \includegraphics[width = 2.5in]{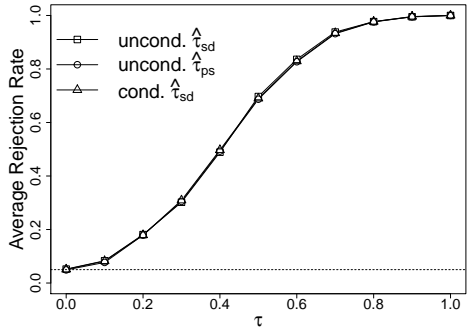}
  \subcaption{$\lambda = 0$}
\end{minipage}
\hspace{0.3in}
\begin{minipage}[b]{0.45\linewidth}
  \includegraphics[width = 2.5in]{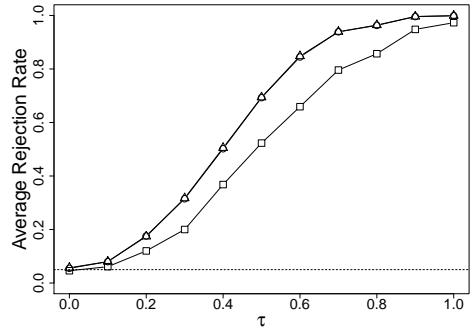}
  \subcaption{$\lambda = 1.5$}
\end{minipage}
\begin{minipage}[b]{0.45\linewidth}
  \includegraphics[width = 2.5in]{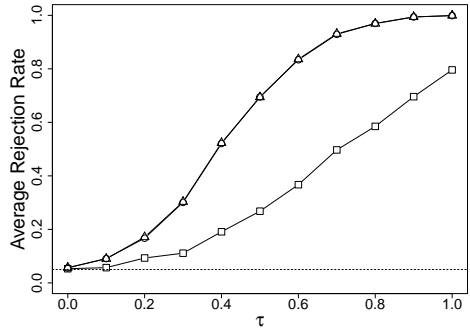}
  \subcaption{$\lambda = 3$}
\end{minipage}
\end{center}
\caption{\textbf{Unconditional average rejection rates for different $\tau$ and $\lambda$}}
\label{uncond_sim}
\end{figure}

Figure \ref{uncond_sim} reports the unconditional rejection rates for different values of $\tau$ and  $\lambda$.  The units were assigned using the completely randomized assignment mechanism.

\noindent When $\lambda = 0$, Figure \ref{uncond_sim}(a), the covariate is not prognostic and the three tests are virtually the same.  All reject the null hypothesis with probability 0.05 (the horizontal dotted line) under the null of $\tau = 0$, and, as expected, the power increases as $\tau$ increases.  In Figure \ref{uncond_sim}(b), the covariate is more prognostic, $\lambda = 1.5$, and the unconditional test using $\hat{\tau}_{ps}$ and the conditional test appear unchanged but the power of the unconditional test using $\hat{\tau}_{sd}$, shown in the black line, falls.  The unconditional test using $\hat{\tau}_{sd}$ is the one test that ignores the covariate balance.  It is more of the same in Figure \ref{uncond_sim}(c), where again the unconditional test using $\hat{\tau}_{ps}$ and the conditional test appear unchanged.  However, the power of the unconditional test using $\hat{\tau}_{sd}$ falls even lower.  In summary, as the covariate becomes more prognostic, the power of the unconditional test using $\hat{\tau}_{sd}$ decreases while the power of the other two tests remain the same.
We should adjust for covariate imbalance either by modifying the test statistic or by conditioning, but little distinguishes between the two approaches.

% ----------------------------------------------------------------
\subsection{Conditional properties}
% ----------------------------------------------------------------

\begin{figure}[ht]
\begin{center}
\begin{minipage}[b]{0.45\linewidth}
  \includegraphics[width = 2.5in]{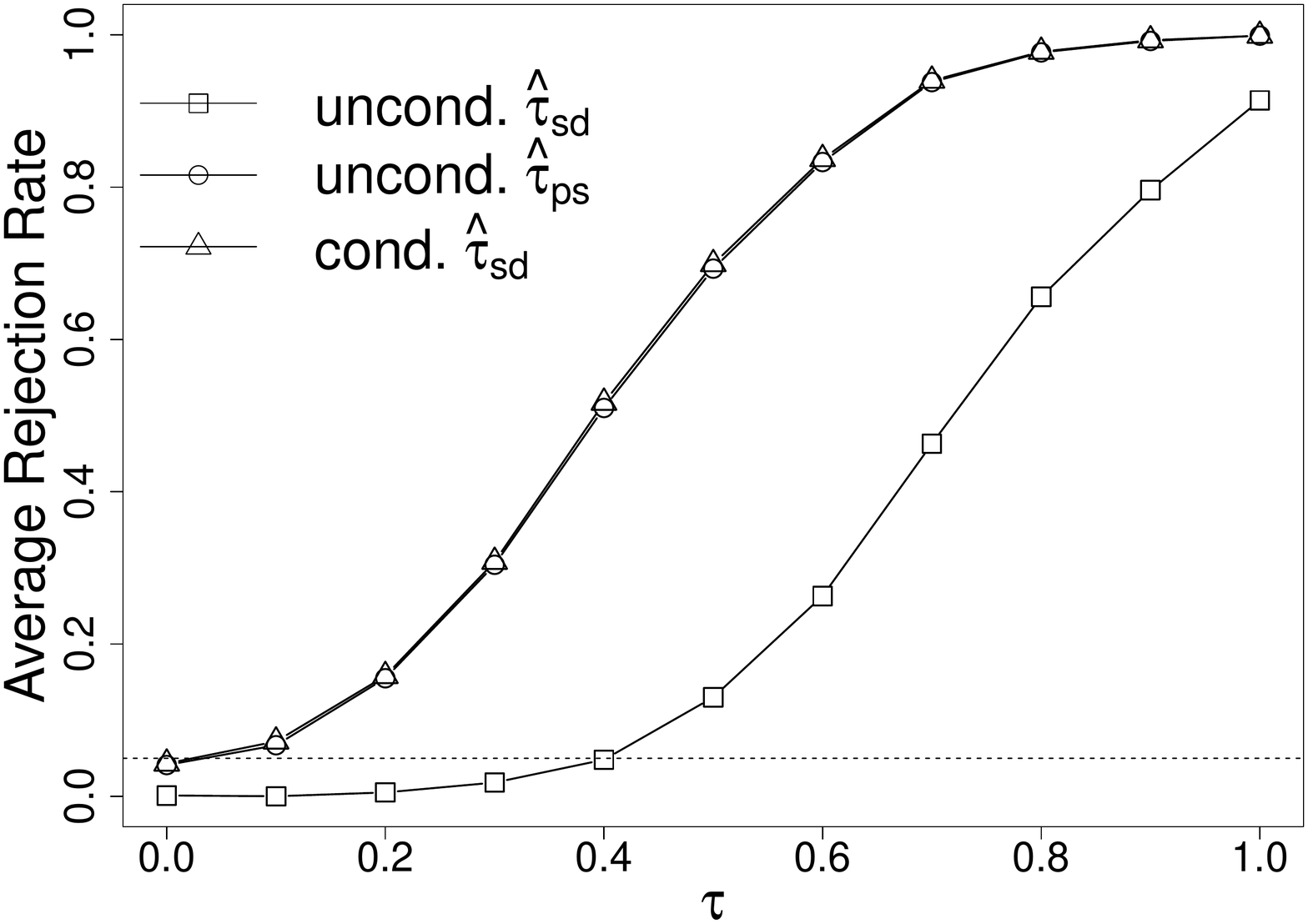}
  \subcaption{$N_{T1} = 25$}
\end{minipage}
\hspace{0.3in}
\begin{minipage}[b]{0.45\linewidth}
  \includegraphics[width = 2.5in]{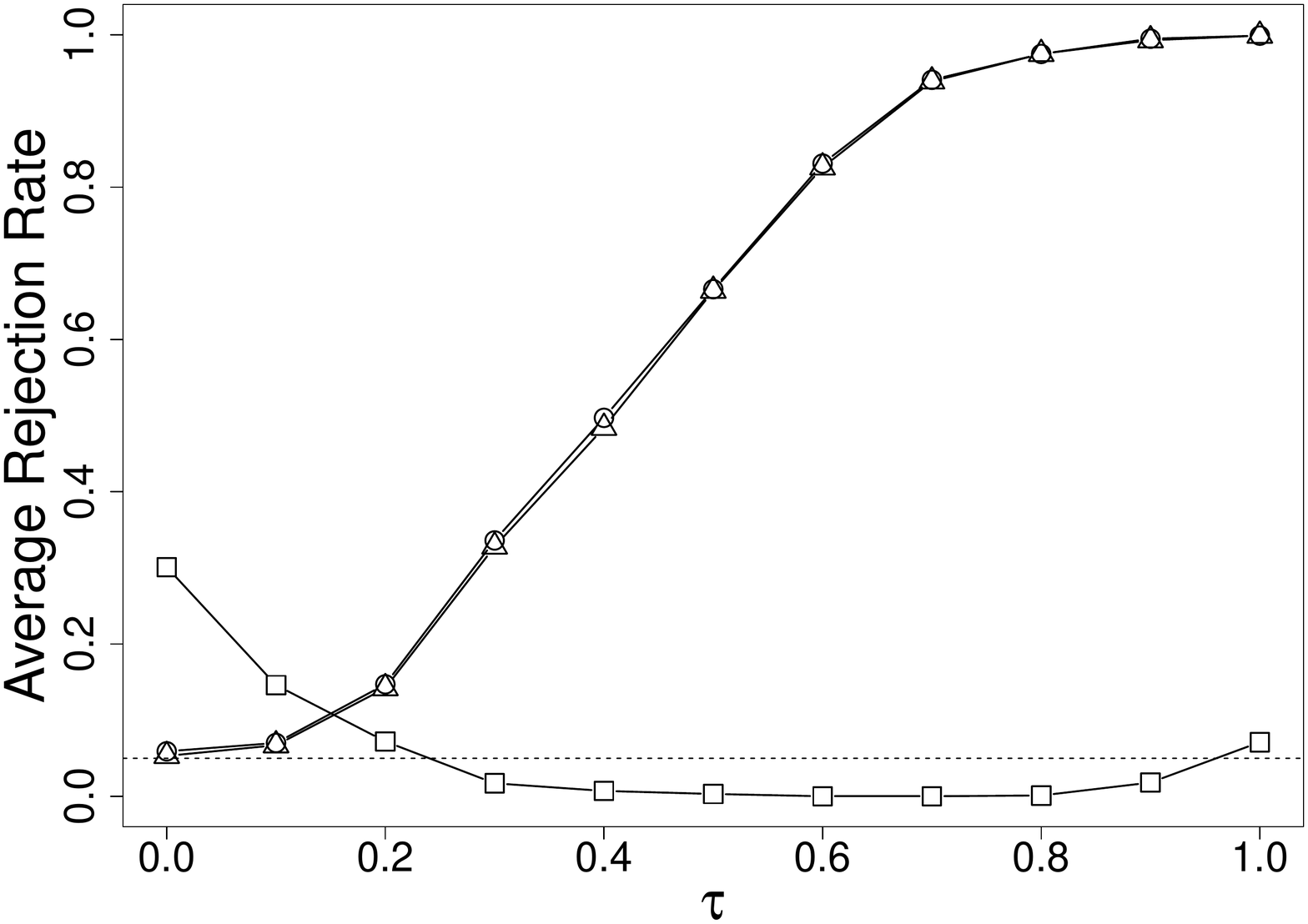}
  \subcaption{$N_{T1} = 30$}
\end{minipage}
\begin{minipage}[b]{0.45\linewidth}
  \includegraphics[width = 2.5in]{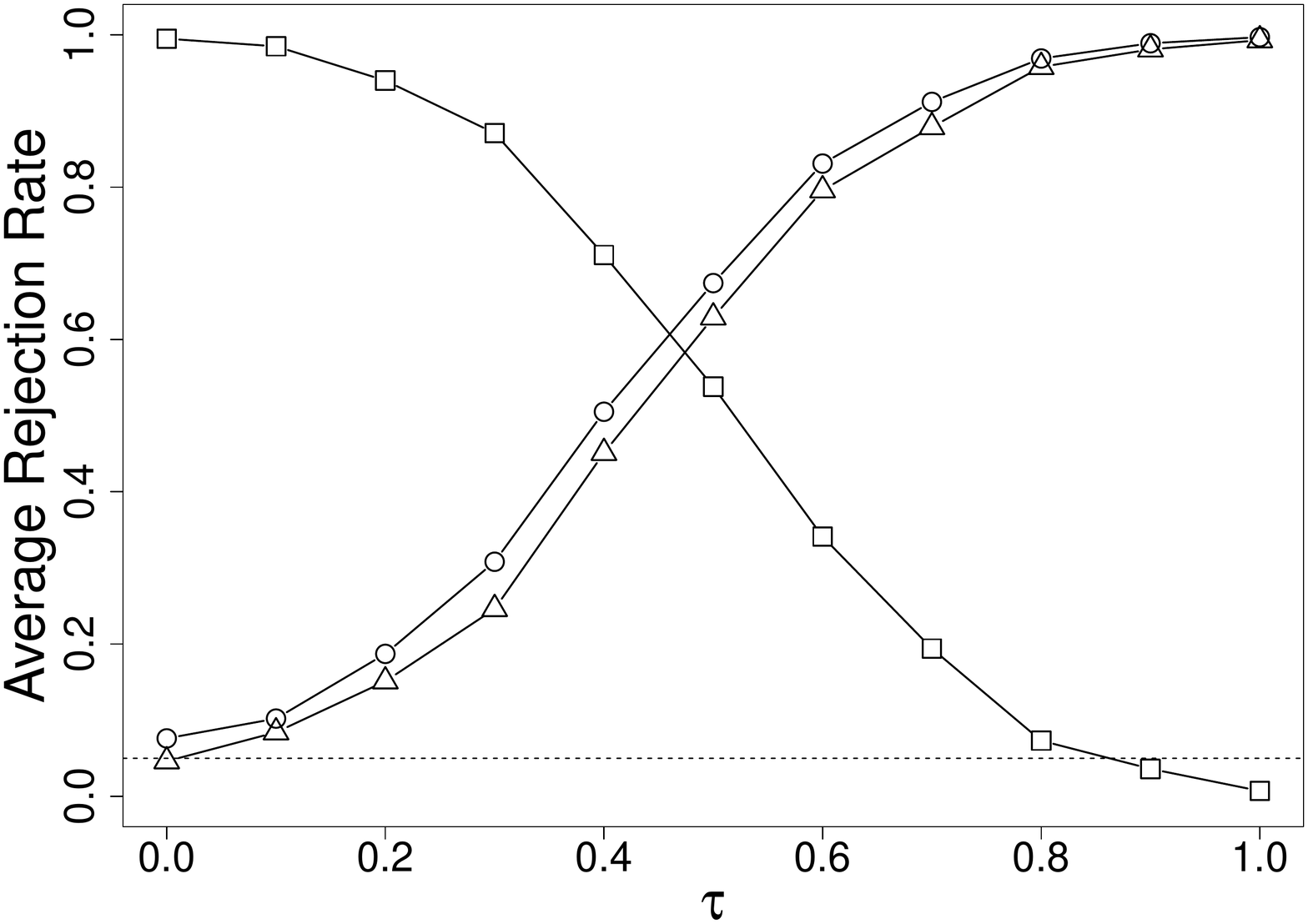}
  \subcaption{$N_{T1} = 35$}
\end{minipage}
\hspace{0.3in}
\begin{minipage}[b]{0.45\linewidth}
  \includegraphics[width = 2.5in]{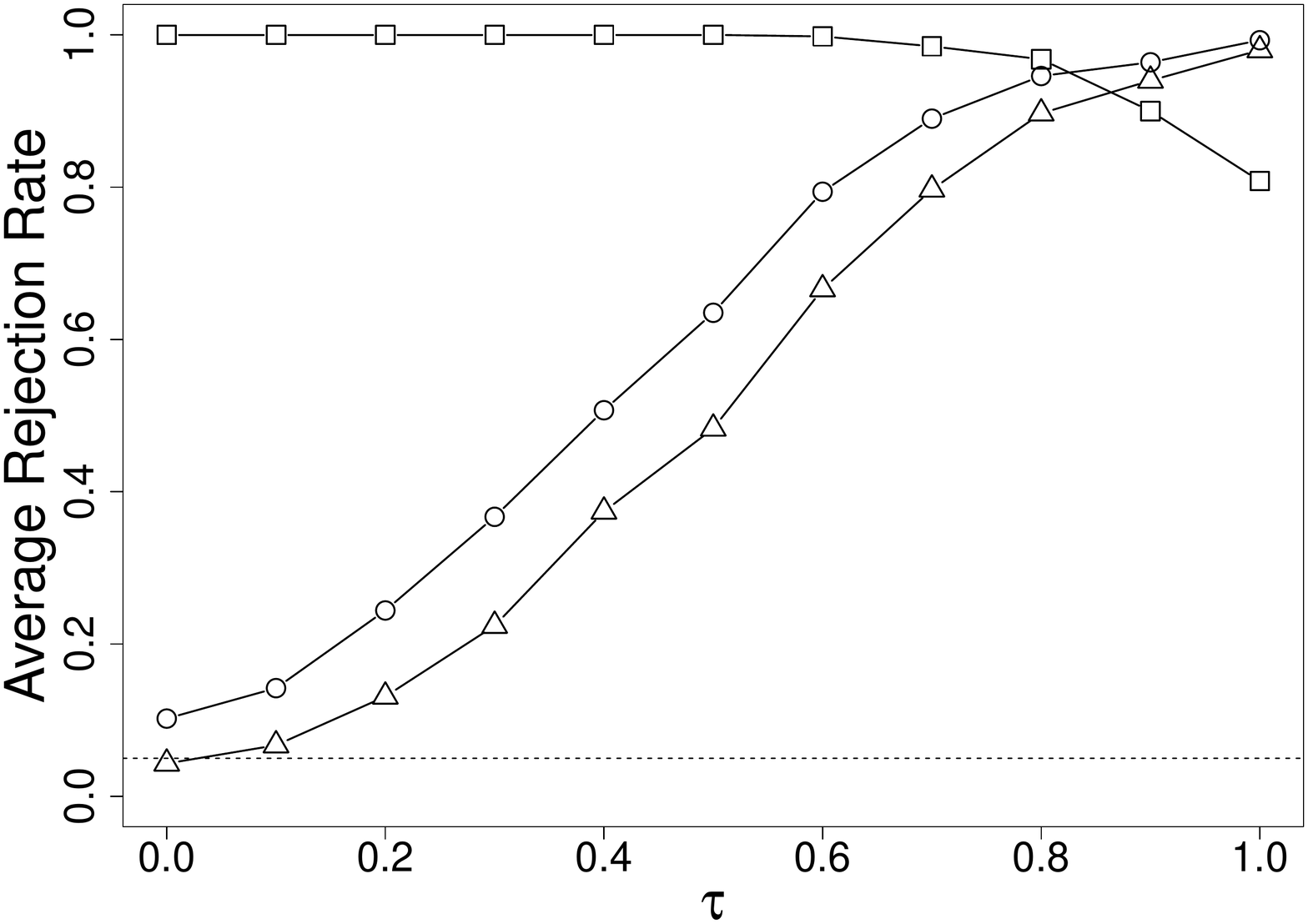}
  \subcaption{$N_{T1} = 40$}
\end{minipage}
\end{center}
\caption{\textbf{Conditional average rejection rates for different $\tau$ and $N_{T1}$:} In all simulations, $\lambda = 3$.}
\label{cond_sim}
\end{figure}

Figure \ref{cond_sim} reports the conditional rejection rates for the three tests under the most prognostic scenario, varying the values of $\tau$ and $N_{T1}$.
In all subfigures $\lambda = 3$.

When $N_{T1} = 25$, Figure \ref{cond_sim}(a), the prognostic covariate is perfectly balanced.  When $\tau = 0$, both the unconditional test using $\hat{\tau}_{ps}$ and the conditional test reject the sharp null with probability 0.05. The unconditional test using $\hat{\tau}_{sd}$ rejects the sharp null with probability less than 0.05.
A simple argument explains this phenomenon: because the covariate is perfectly balanced, $E(\hat{\tau}_{sd} \, | \, N_{T1} = 25) = 0$, the value of $\tau$.
The unconditional randomization test using $\hat{\tau}_{sd}$ compares the test statistic to a reference distribution centered at 0 with variance $\text{var}(\hat{\tau}_{sd})$; however, conditioned on $N_{T1} = 25$, the observed test statistics have a smaller actual variance.
I.e., $\text{var}(\hat{\tau}_{sd}) > \text{var}(\hat{\tau}_{sd} \, | \, N_{T1} = 25)$ because the covariate is prognostic.
Because of this, the test statistics rarely end up in the tails of the reference distribution and the rejection rate is less than 0.05.

As we move from perfect covariate balance to covariate imbalance, Figure \ref{cond_sim}(b), the unconditional test using $\hat{\tau}_{ps}$ and the conditional test appear unchanged, but the unconditional test using $\hat{\tau}_{sd}$ begins to break down.  When $\tau = 0$, $E(\hat{\tau}_{sd}) = 0$ but because the covariate is prognostic, $E(\hat{\tau}_{sd} \, | \, N_{T1} = 30) < 0$.  Thus, the unconditional test is comparing the observed test statistics, which tend to be negative, to a reference distribution centered at 0.  As seen in Figure \ref{cond_sim}(b), this gives a rejection rate greater than 0.05 when $\tau = 0$.  As $\tau$ increases, $E(\hat{\tau}_{sd} \, | \, N_{T1} = 30)$ increases since the positive treatment effect counteracts the effect of the covariate imbalance.  Thus, the observed test statistics are pushed closer to 0 and the rejection rate falls.  Eventually, the treatment effect overcomes the covariate imbalance and the rejection rate begins to rise, which we see at $\tau = 1$.

In Figures \ref{cond_sim}(c) and \ref{cond_sim}(d), as the covariate imbalance increases, the unconditional test using $\hat{\tau}_{sd}$ repeats this pattern.  More interestingly, as the covariate imbalance increases, we also begin to see differences between the unconditional test using $\hat{\tau}_{ps}$ and the conditional test.
In Figure \ref{cond_sim}(d), for example, the unconditional test using $\hat{\tau}_{ps}$ rejects the sharp null with probability over 0.05 when $\tau = 0$: the test has the wrong conditional significance level.  In contrast, although the power of the conditional test has dropped slightly, its conditional significance level is still 0.05.  The key to understanding why the conditional significance level is incorrect for the unconditional test using $\hat{\tau}_{ps}$ is that the conditional variance of $\hat{\tau}_{ps}$ increases with the covariate imbalance.  Thus, $\text{var}(\hat{\tau}_{ps} \, | \, N_{T1} = 40) > \text{var}(\hat{\tau}_{ps})$ and the observed test statistics are more spread out than the reference distribution they are being compared to.

The unconditional properties supported the notion that we should adjust for covariate imbalance either by modifying the test statistic or by conditioning.  The conditional properties indicate that modifying the test statistic is inferior to conditioning because unconditional tests with modified test statistics can still have the wrong conditional signficance level.

% ----------------------------------------------------------------
\section{Product marketing example}
% ----------------------------------------------------------------

Our product marketing experiment involved roughly $2000$ experimental subjects and $K = 11$ treatment levels, which were eleven versions of a particular product.  Each subject randomly received by mail one of the products.  Each subject used the product and returned a survey regarding the product's performance.  The outcome of interest was an ordinal variable with three levels, 1, 2, and 3 (with 3 being the best), and the goal was to identify which product version the subjects preferred.  The survey also collected covariate information, such as income and ethnicity, and the experimenters were concerned about the effect of covariate imbalance on their conclusions.  Critically, covariate information was not collected until after the units were assigned to treatment and thus blocking and rerandomization were not possible.

After removing observations with missing values under the assumption that missingness is not related to the product, there were $N = 2256$ experimental units.  The number of units assigned to each treatment level is given on Table \ref{prop_table}.

\begin{table}[H]
\caption{\textbf{Number of units assigned to each treatment level:} The number of units assigned to each treatment level was relatively equal.  }
\begin{center}
{\small
\begin{tabular}{l|ccccccccccc}
\multicolumn{1}{c}{} & \multicolumn{11}{c}{Treatment} \\
& 1 & 2 & 3 & 4 & 5 & 6 & 7 & 8 & 9 & 10 & 11 \\
\hline
\# of Units & 238 & 266 & 225 & 231 & 237 & 226 & 198 & 135 & 136 & 136 & 228 \\
Percentage & 10\% & 12\% & 10\% & 10\% & 11\% & 10\% &  9\% &  6\% &  6\% &  6\% & 10\%
\end{tabular}
}
\end{center}
\label{prop_table}
\end{table}

We first conduct an omnibus test and then a set of pairwise tests. In the omnibus test, we test the sharp null hypothesis that all $K$ unit level potential outcomes are equal:
\begin{equation}
	 H_0: Y_i(1) = \cdots = Y_i(11) \mbox{ for all } i = 1, \ldots, N . \label{eq:omnibus_null}
\end{equation}
If we reject the sharp null, we move on to the pairwise tests, where we compare all ${11 \choose 2} = 55$ pairs of treatments to rank the products.

For the omnibus test, we use the Kruskal-Wallis statistic as the test statistic \citep{Kruskal1952}. This statistic is typically used in the Kruskal-Wallis test, a non-parametric test similar to one-way ANOVA, and is similar to the $F$-statistic in that it is a ratio of sum of squares. Larger values of the statistic indicate that the treatment levels are different. The test statistic is given by
\begin{equation}
  (N - 1) \frac{\sum_{j=1}^K N_j (\bar{r}_j^{\text{obs}} - \bar{r}^{\text{obs}})^2}{\sum_{i = 1}^N (r_i^{\text{obs}} - \bar{r}^{\text{obs}})^2 }, \label{eq:kruskal}
\end{equation}
where $\bar{r}^{\text{obs}}_j$ is mean rank in the $j$th treatment level and $\bar{r}^{\text{obs}}$ is the mean rank overall. In our example, the response is ordinal, and thus we can directly use the observed data $y$ instead of the ranks $r$ in (\ref{eq:kruskal}).

For the pairwise test, we use the difference of the mean ranks as the test statistic. While testing the difference between treatment groups $j$ and $\tilde{j}$, we use observed outcomes only from those units that are assigned either to treatment $j$ or $\tilde{j}$.

To explore the difference between the conditional and unconditional tests, we first analyze the data from the unconditional perspective, and then re-analyze the same data conditioning on blocks formed out of covariates. For both the omnibus and pairwise tests, the randomization distributions of the test statistics were obtained from 1000 permutations in each case. We first report the results of the unconditional versions of the omnibus and pairwise tests, followed by the conditional tests. We do not consider adjustments for multiple testing, because that is not the focus of this paper. To account for multiple testing, one can use simple but conservative methods like Bonferroni correction or methods that control the False discovery rate, but the procedures proposed in this paper remain exactly the same.

\subsection{Unconditional test}

\begin{figure}[H]
  \begin{center}
    \includegraphics[width=3.2in]{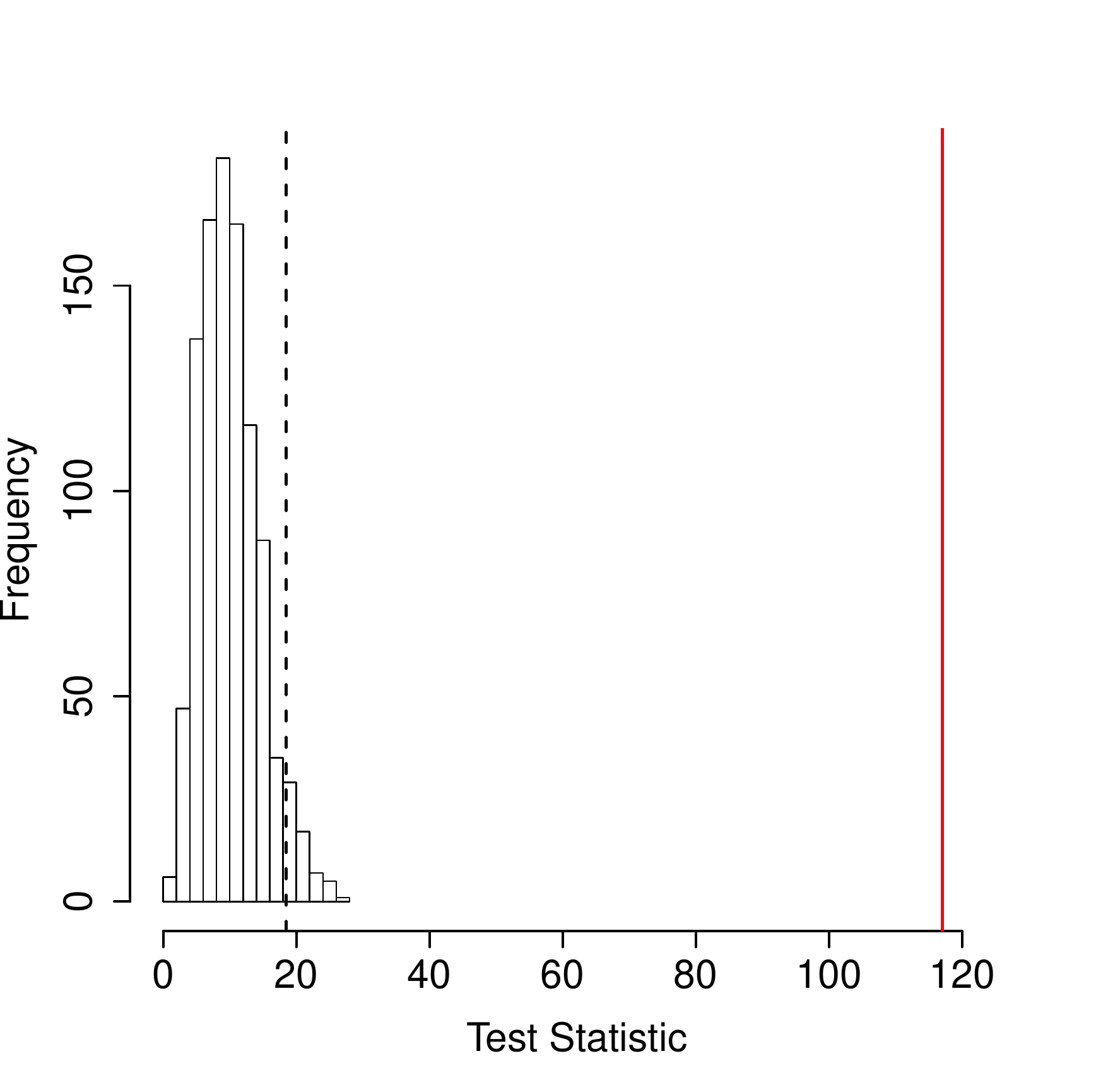}
  \end{center}
  \caption{\textbf{Unconditional randomization test using Kruskal-Wallis test statistic} }
  \label{omnibus_uncond}
\end{figure}

The results of the unconditional omnibus test using the Kruskal-Wallis statistic is shown in Figure \ref{omnibus_uncond}, in which the vertical red line is the observed value of the test statistic and the dashed line is the 95th quantile of the reference distribution.  The histogram is the (unconditional) distribution of the test statistic under the sharp null hypothesis. The observed test statistic is 18.92, and the $p$-value is approximately zero. Thus there is a very strong evidence that the products are different.

\begin{table}[ht]
\caption{The $p$-values for unconditional pairwise tests} \label{t:pairwise_uncond}
\centering
\footnotesize
\begin{tabular}{r|rrrrrrrrrrr}
  \hline
& 1 & 2 & 5 & 4 & 11 & 3 & 6 & 8 & 10 & 9 & 7 \\
  \hline
1 &  & 0.06 & 0.03 & 0.01 & 0.00 & 0.00 & 0.00 & 0.00 & 0.00 & 0.00 & 0.00 \\
2 &  &  & 0.56 & 0.43 & 0.00 & 0.00 & 0.00 & 0.00 & 0.00 & 0.00 & 0.00 \\
5 &   &  &  & 0.85 & 0.01 & 0.01 & 0.02 & 0.00 & 0.02 & 0.00 & 0.00 \\
4 &   &  &  &  & 0.02 & 0.03 & 0.03 & 0.02 & 0.02 & 0.00 & 0.00 \\
11 &   &  &  &  &  & 1.00 & 1.00 & 0.70 & 0.62 & 0.16 & 0.00 \\
3 &   &  &  &  &  &  & 0.98 & 0.70 & 0.65 & 0.20 & 0.00 \\
6 &   &  &  &  &  &  &  & 0.76 & 0.64 & 0.23 & 0.00 \\
8 &   &  &  &  &  &  &  &  & 0.90 & 0.39 & 0.00 \\
10 &   &  &  &  &  &  &  &  &  & 0.54 & 0.00 \\
9 &   &  &  &  &  &  &  &  &  &  & 0.01 \\
7 &   &  &  &  &  &  &  &  &  &  &  \\
   \hline
\end{tabular}
\end{table}

The results of the pairwise tests are summarized in Table \ref{t:pairwise_uncond}, in which treatments are arranged in descending order with respect to their average outcomes (treatment 1 has the largest average whereas treatment 7 has the smallest). From Table \ref{t:pairwise_uncond} we observe that treatment 1 appears to be the most favored one, although the difference between treatments 1 and 2 is not statistically significant at level 0.05. Next, we perform the conditional test to check if these two treatments can be separated further by conditioning on the observed covariate distribution.

\subsection{Conditional tests}

For this analysis, we consider the following eight covariates, all of which are categorical: (i)  order of detergent (3 levels), (ii) under stream (2 levels), (iii) care for dishes (5 levels), (iv) water hardness (5 levels), (v) consumer segment (4 levels), (vi) household income (11 levels), (vii) age (6 levels) and (viii) hispanic (2 levels). This gives $3 \times 2 \times 5 \times 5 \times 4 \times 11 \times 6 \times 2 = 79200$ different unique combinations of our covariates. To reduce the number of potential categories, we then cluster the observations based on these covariates (but not outcomes or treatment assignment) to create a new pre-treatment categorical covariate that we can condition on. We consider clustering a simple but useful first step in carrying out a conditional randomization test. The advantage of the clustering method is that we can replace the eight categorical covariates with one categorical covariate, the cluster indicator.

Because the covariates are categorical, we use the $k$-modes algorithm introduced by \citet{Huang1997a}, which extends the $k$-means algorithm to  handle categorical variables. Details of this step are described in Appendix B, in which we make an attempt to identify the correct number of clusters using an elbow plot shown in Figure \ref{elbow}. It appears from the plot that choosing the optimum number of clusters as seven is a reasonable choice. Table \ref{cluster_tab} shows the two-way distribution of experimental units over the seven clusters and assigned treatments.

\begin{table}[ht]
\caption{\small \textbf{Clusters and treatment levels:}  The rows are the seven clusters and the columns are the eleven treatment levels.  Entries are counts of subjects in that cluster given that product.}
\footnotesize
\begin{center}
\begin{tabular}{r|rrrrrrrrrrr|r}
 \multicolumn{1}{c}{} & \multicolumn{11}{c}{Treatment} & \\
 & 1 & 2 & 3 & 4 & 5 & 6 & 7 & 8 & 9 & 10 & 11 &  \\
  \hline
1 & 82 & 93 & 63 & 88 & 83 & 84 & 71 & 39 & 56 & 46 & 78 & 783 \\
  2 & 35 & 28 & 29 & 26 & 28 & 25 & 21 & 13 & 12 & 16 & 27 & 260 \\
  3 & 44 & 37 & 41 & 47 & 34 & 37 & 30 & 32 & 22 & 23 & 44 & 391 \\
  4 & 21 & 29 & 28 & 18 & 22 & 20 & 10 & 17 & 12 & 13 & 21 & 211 \\
  5 & 14 & 26 & 20 & 20 & 18 & 18 & 14 & 8 & 9 & 11 & 22 & 180\\
  6 & 16 & 17 & 22 & 13 & 24 & 17 & 24 & 11 & 10 & 11 & 11 & 176\\
  7 & 26 & 36 & 22 & 19 & 28 & 25 & 28 & 15 & 15 & 16 & 25 & 255 \\
\end{tabular}
\end{center}
\label{cluster_tab}
\end{table}

We then carry out the conditional randomization test by conditioning on the number of units in each cluster assigned to each treatment level.  The result of the omnibus test is similar to that of the unconditional test, and the $p$-value is approximately zero. We next perform the pairwise conditional test, and the results are summarized in Table \ref{t:pairwise_cond}. Comparing the $p$-values in Tables \ref{t:pairwise_uncond} and \ref{t:pairwise_cond}, it appears that the conditional test provides us with a marginally stronger evidence that treatment 1 is better than treatment 2. We thus conclude that product 1 is the most preferred product and that versions 1, 2, 5, and 4 are clearly preferred to the seven other products. Product 7 is definitively the worst. Note that in this example, the improvement achieved by conditioning is marginal. A plausible explanation is, the covariates were actually not as prognostic as they were believed to be.

%\begin{figure}[H]
%  \begin{center}
%    \includegraphics[width=3.5in]{figures/cond_rand_omnibus_cluster.pdf}
% \end{center}
%  \caption{\textbf{Conditional randomization test using Kruskal-Wallis test statistic:} }
%  \label{omnibus}
%\end{figure}

%\begin{figure}[H]
%  \begin{center}
%    \includegraphics[width=3.5in]{figures/avg_ratings.pdf}
%  \end{center}
%\caption{\textbf{Mean outcome by treatment level:} The vertical lines mark the 95\% confidence intervals.}
%\label{mean_outcome_fig}
%\end{figure}

\begin{table}[ht]
\caption{The $p$-values for pairwise conditional tests} \label{t:pairwise_cond}
\centering
\footnotesize
\begin{tabular}{r|rrrrrrrrrrr}
  \hline
& 1 & 2 & 5 & 4 & 11 & 3 & 6 & 8 & 10 & 9 & 7 \\
  \hline
1 &  & 0.04 & 0.02 & 0.01 & 0.00 & 0.00 & 0.00 & 0.00 & 0.00 & 0.00 & 0.00 \\
2 &   &  & 0.55 & 0.47 & 0.00 & 0.00 & 0.01 & 0.00 & 0.01 & 0.00 & 0.00 \\
5 &   &  &  & 0.88 & 0.01 & 0.02 & 0.02 & 0.02 & 0.02 & 0.00 & 0.00 \\
4 &   &  &  &  & 0.02 & 0.02 & 0.02 & 0.01 & 0.04 & 0.00 & 0.00 \\
11 &   &  &  &  &  & 0.94 & 0.98 & 0.64 & 0.65 & 0.22 & 0.00 \\
3 &   &  &  &  &  &  & 0.98 & 0.88 & 0.74 & 0.17 & 0.00 \\
6 &   &  &  &  &  &  &  & 0.68 & 0.70 & 0.29 & 0.00 \\
8 &   &  &  &  &  &  &  &  & 0.92 & 0.46 & 0.00 \\
10 &   &  &  &  &  &  &  &  &  & 0.52 & 0.00 \\
9 &    &  &  &  &  &  &  &  &  &  & 0.01 \\
7 &   &  &  &  &  &  &  &  &  &  &  \\
   \hline
\end{tabular}
\end{table}

% ----------------------------------------------------------------
\section{Conclusion}
% ----------------------------------------------------------------

We considered conditional randomization tests as a form of covariate adjustment for randomized experiments.  Conditional randomization tests have received relatively little attention in the statistics literature and we built upon \citet{Rosenbaum1984} and \citet{Zheng2008} by introducing original notation to prove that the conditional randomization test has the correct unconditional significance level and to describe covariate balance more formally.  Our simulation results verify that conditional randomization tests behave like more traditional forms of covariate adjustment but have the added benefit of having the correct conditional significance level.

%\luke{Put the associated discussion back into paper?}

The conditional randomization test conditioning on the observed covariate balance shares similarities with rerandomization \citep{Morgan2012}.  Rerandomization is a treatment assignment mechanism that restricts $\mathscr{S}$ to the set of treatment assignments which satisfy a pre-determined level of covariate balance.  A balance criterion, $B(\bm{w}, \textbf{X})$, determines if the treatment assignment is acceptable,  $B(\bm{w}, \textbf{X}) = 1$, or unacceptable, $B(\bm{w}, \textbf{X}) = 0$.  Thus, $\mathscr{S} =  \lbrace \omega \, : \, B(\omega, \textbf{X}) = 1 \rbrace$.  As a result, the observed treatment assignment is guaranteed to be balanced on covariates. The experiment is then analyzed using a randomization test where the reference set is $\mathscr{S}$.
%Rerandomization is similar to restriced randomization but what makes rerandomization unique is that it is designed to balance multiple covariates simultaneously.

The conditional randomization test is like a \textit{post-hoc rerandomization test}.  In a conditional randomization test, we observe some treatment assignment, $\bm{w} = \bm{w}$, and covariate balance, $B(\bm{w}, \textbf{X}) = b$, and then act as if that treatment assignment were drawn from some partition with the same covariate balance, $\mathscr{S}_{b}$.  The rerandomization test and conditional randomization test would be identical if, for instance, $\mathscr{S}_{b} = \lbrace \omega \, : \, B(\omega, \textbf{X}) = 1 \rbrace$.
Both methods allow for balancing multiple covariates simultaneously.

As pointed out by a reviewer, the proposed approach has benefits in both ``unlucky'' and ``lucky'' randomizations. For an ``unlucky'' randomization, it will adjust the null distribution to account for covariate imbalance, working to preserve Type I error in a conditional sense. For a ``lucky'' randomization, it will restrict the tails of the null distribution increasing power.

One limitation of conditional randomization tests is, drawing randomizations from a partition can be computationally expensive, if done with simple re-sampling and acceptance/rejection approaches. For a single categorical covariate, we can sample more directly. However, for multiple categorical covariates where we control all of the margins, this becomes more difficult. Thus, one area of future research is exploration of sampling techniques using different types of covariate balance functions. Whereas clustering appears to be a useful first step, balance functions that take into account the joint distributions of covariates and thus have a tensor structure may practically be more meaningful. However, sampling from reference sets based on such balance functions can be quite challenging and requires further investigation.

\vspace{0.2 in}

\noindent \textbf{Acknowledgement}: We are grateful to two reviewers for their insightful comments that resulted in substantial improvements in the contents and the presentation of the paper.

\section*{Appendix A}

We here prove that tests using $\hat{\tau}_{ps}$ are equivalent to tests using $\hat{\tau}_{sd}$ when conditioning on balance of a categorical covariate.

First note that $\hat{\tau}_{ps} = \hat{\beta}_W$, where $\hat{\beta}_W$ is the estimate of $\beta_W$ from the linear regression with interactions between $X$ and $W$:
\begin{equation}
  Y_i^{\text{obs}} = \beta_0 + \beta_W W_i + \sum_{k = 2}^K \beta_k X_{ik} + \sum_{k=2}^K \gamma_{k} (W_i \cdot X_{ik}) + \epsilon_i
\end{equation}

\noindent where

\begin{equation}
   X_{ik} = \left\{
     \begin{array}{rl}
       1 & : \text{if the $i$th unit is in the $k$th stratum}\\
       -1 & : \text{if the $i$th unit is in the first stratum} \\
       0 & : \text{otherwise.}
     \end{array}
   \right.
\end{equation}

The dummies $X_i$ follow the sum contrast coding.
We next show that, conditioning on the observed balance, $\hat{\tau}_{ps}$ is a monotonic function of $\hat{\tau}_{sd}$.

Let $\lbrack \bm{w},\, \textbf{F}\rbrack$ denote the design matrix, where \textbf{F} includes a column of ones for the intercept and columns for the categorical indicator variables and interactions.  Also, note that $\bm{w}^T \yobs = (\hat{\tau}_{sd} + \frac{1}{N_C} \textbf{1}^T \yobs) / (\frac{1}{N_T} + \frac{1}{N_C})$.
%\luke{Why is this?  Explain further.}
  Let $P_\textbf{F} = \textbf{F} (\textbf{F}^T \textbf{F})^{-1} \textbf{F}^T$ be the projection matrix onto the columns of $\textbf{F}$.  We then use the regression anatomy formula \citep{Angrist2009}.

\begin{equation*}
  \hat{\tau}_{ps} = \hat{\beta}_W = \frac{\bm{w}^T (I - P_{\textbf{F}}) \yobs}{\bm{w}^T (I - P_{\textbf{F}}) \bm{w}}.
\end{equation*}

\noindent Note that conditioning on the observed balance implies that $\bm{w}^T \textbf{F}$ is a constant and thus

\begin{equation*}
  \hat{\tau}_{ps} = \frac{\bm{w}^T \yobs - k_1}{k_2} = \left( \frac{1}{k_2} \right) \frac{\hat{\tau}_{sd} + \frac{1}{N_C} 1^T\Yobs}{ \frac{1}{N} + \frac{1}{N_C}} - \frac{k_1}{k_2}
\end{equation*}

\noindent where $k_1 = \bm{w}^T P_{\textbf{F}} \yobs$ and $k_2 = \bm{w}^T (I - P_{\textbf{F}}) \bm{w}$.  Finally, since $\bm{w}^T \yobs$ is a monotonic function of $\hat{\tau}_{sd}$, $\hat{\tau}_{ps}$ is also a monotonic linear function of $\hat{\tau}_{sd}$.

Finally, because $\widehat{\tau}_{ps}$ is a monotone scaling of $\widehat{\tau}_{sd}$, $\pr{ \widehat{\tau}_{ps} > t^{obs}_{ps} } = \pr{ \widehat{\tau}_{sd} > t^{obs}_{sd} }$ under the null since the rejection region for the post-stratified estimator is merely the rescaled rejection region for the simple-difference estimator.
I.e., any potential randomization $\bm{w}$ will result in an equivalently ``extreme'' $t$, as defined by its quantile, regardless of choice of statistic.

\section*{Appendix B}
We used $k$-modes to collapse many categorical covariates into a few groups to allow for easier conditional randomization.
The $k$-modes algorithm relies on a dissimilarity measure, $d(\cdot, \cdot)$, which measures the dissimilarity between two observations.  The dissimilarity measure is the number of categorical variables which are different between the two obervations.  So, if $X_i = (1, 2, 4, 2, 1, 10, 3, 1)$ and $X_j = (2, 1, 4, 2, 1, 10, 3, 1)$, then $d(X_i, X_j) = 2$.  The smaller the dissimilarity measure the more similar the two observations.  This is a simple measure: it gives equal weight to all covariates and completely ignores the ordinal structure of some of the categorical variables.  For instance, an income value of 11 is much closer to an income value of 10 than to 1 but this aspect is ignored here.
Other dissimilarity measure are certainly possible.  The mode of a set of observations, $\lbrace X_1, \ldots, X_n \rbrace$, is the vector $Q$ that minimizes

\begin{equation}
  \sum_{i = 1}^{n} d(Q, X_i).
\end{equation}

\begin{figure}[H]
  \begin{center}
    \includegraphics[width=3.5in, angle=90]{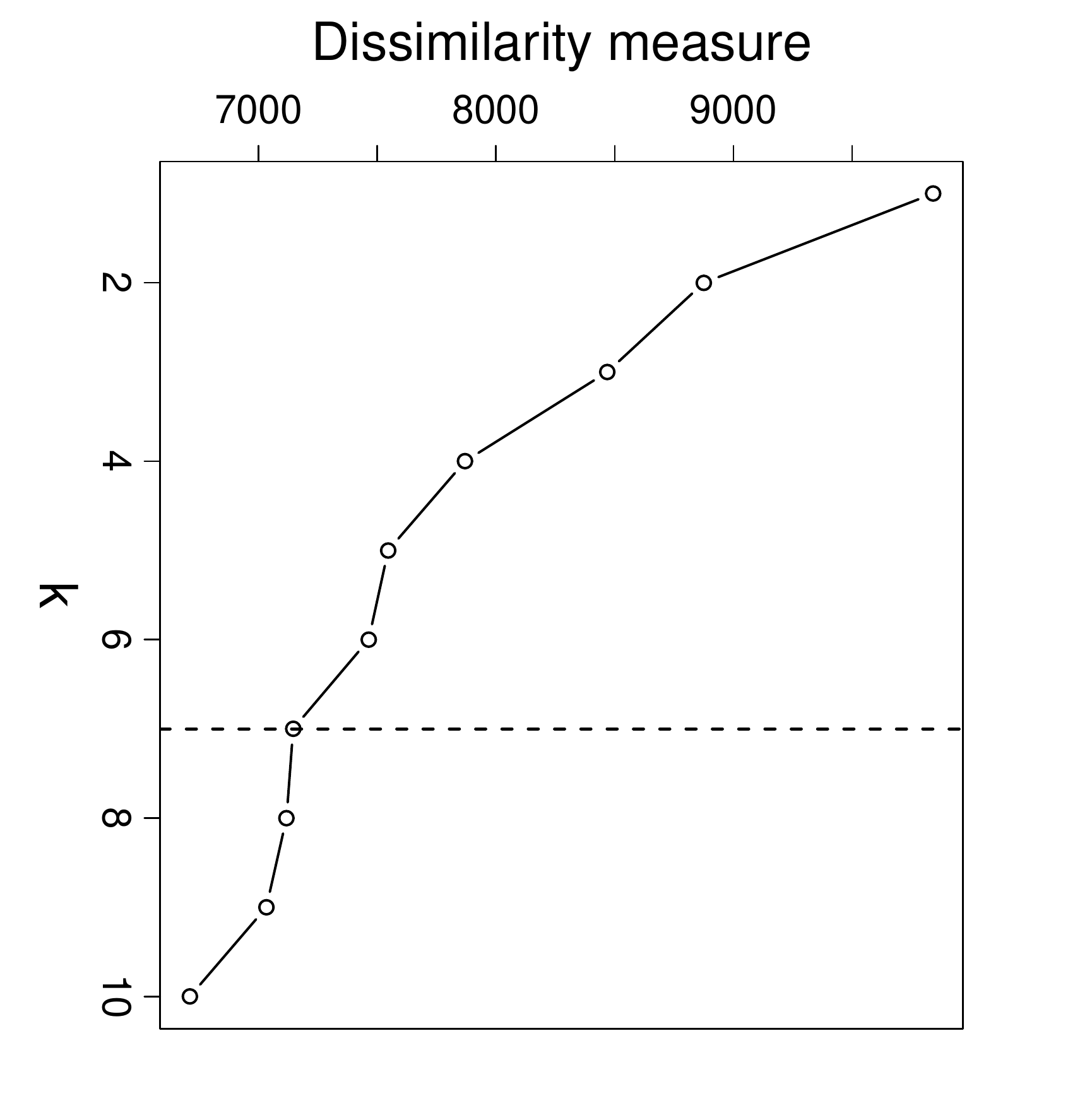}
  \end{center}
  \caption{\textbf{Elbow plot:}  The elbow is determined at $k = 7$, the vertical dashed line.}
  \label{elbow}
\end{figure}

\noindent The $k$-modes algorithm follows the familar steps of the $k$-means algorithm: Start with $k$ candidate modes.
Then assign each observation to the closest mode according to the dissimilarity measure.
Finally re-calculate the modes of each cluster and repeat these last two steps until convergence.
We  determined an appropriate number of clusters, $k$, via an elbow plot, shown in Figure \ref{elbow}.

\noindent In this case, $k = 7$ seems to be a reasonable choice.  The contingency table, in Table \ref{cluster_tab}, summarizes the number of units in each cluster assigned to each treatment level.

% Correction
\newpage
\section*{Erratum}

There was an error in this paper and in \citet{Hennessy2016} and we are very grateful to Peng Ding for pointing it out.  

Proposition 1, restated below, is incorrect, our proof being a mis-application of a result from \citet{Rosenbaum1984}.

\bigskip

\textbf{Proposition 1.} Let $X$ denote a categorical covariate with $J$ levels, observed after a two-armed randomized experiment is conducted with $N$ units. Let $N_j$ denote the observed number of units that belong to stratum $j$, and let $N_{Tj}$ and $N_{Cj}$ denote the number of units assigned to treatment and control respectively, in stratum $j$, such that $N_{Tj} + N_{Cj} = N_j$, and $\sum_{j=1}^J N_j = N$. Then the conditional randomization test using the simple difference test statistic $\hat{\tau}_{sd} = \bar{Y}^{\text{obs}}_T - \bar{Y}^{\text{obs}}_C$ and the balance function ($N_{T1}$, ..., $N_{TJ}$) is equivalent to the conditional randomization test using the composite test statistic $\hat{\tau}_{ps} = \sum_{j=1}^J \frac{N_j}{N} \hat{\tau}_{sd, j}$, where $\hat{\tau}_{sd, j}$ denotes the simple difference test statistic for the $j$th stratum.

\bigskip

We can show the proposition is not true by a simple counterexample.  In order for the two conditional tests to be equivalent, they must yield the same p-values. Consider the situation where $N = 5$, $X = (1, 1, 1, 2, 2)$, $\textbf{w} = (1, 0, 0, 1, 0)$, and $\textbf{y}^{\text{obs}} = (1.13, 0.49, -0.31, 0.98, 1.68)$.  In this case, $\hat{\tau}_{sd} = 0.435$ and $\hat{\tau}_{ps} = 0.344$.  To find the p-values for the conditional test, we consider the values of the test statistics across all 6 alternative randomizations where $N_{T1} = 1$ and $N_{T2} = 1$.  

\begin{table}[H]
\caption{\textbf{Alternative randomizations:} For each alternative randomization where $N_{T1} = 1$ and $N_{T2} = 1$, we calculate both test statistics.}
\begin{center}
  \begin{tabular}{l|r|r}
Randomization & $\hat{\tau}_{sd}$ & $\hat{\tau}_{ps}$  \\
\hline
(1, 0, 0, 1, 0) & 0.435 & 0.344 \\
(0, 1, 0, 1, 0) & -0.098 & -0.232 \\
(0, 0, 1, 1, 0) & -0.765 & -0.952 \\
(1, 0, 0, 0, 1) & 1.018 & 0.904 \\
(0, 1, 0, 0, 1) & 0.485 & 0.328 \\
(0, 0, 1, 0, 1) & -0.182 & -0.392 \\
\end{tabular}
\end{center}
\label{cont_table}
\end{table}

\noindent To calculate the p-values, we find the proportion of test statistics as or more extreme than the observed.  For $\hat{\tau}_{sd} = 0.435$, there are three test statistics as large or larger (0.435, 0.485, 1.018), so the 2-sided p-value is $2 \cdot 3/6 = 1$.  For $\hat{\tau}_{ps} = 0.344$, there are two test statistics as large or larger (0.344, 0.904), so the 2-sided p-value is $2 \cdot 2/6 = 2/3$.  Since the p-values do not agree, the tests are not equivalent. 

The incorrect proof in Appendix A mis-applied a result from \citet{Rosenbaum1984} that showed that in a linear regression of the response on the treatment indicator and covariates, a conditional randomization test based on the treatment indicator coefficient is equivalent to the conditional randomization test based on the simple difference test statistic.  This proof rests on the fact that the columns corresponding to the covariates are fixed across randomizations.  While our $\hat{\tau}_{ps}$ does equal a regression coefficient, that regression includes interactions between the treatment indicator and the covariates.  However, these interaction terms are not fixed across the different randomizations and we ignored this fact in the proof.  In the proof in Appendix A, we incorrectly assumed $k_1 =  \textbf{w}^T \textbf{F} (\textbf{F}^T \textbf{F})^{-1} \textbf{F}^T \textbf{y}^{\text{obs}}$ is a constant.  While $\textbf{w}^T \textbf{F} (\textbf{F}^T \textbf{F})^{-1}$ is a constant, $\textbf{F}^T \textbf{y}^{\text{obs}}$ is not.\footnote{We also incorrectly defined the covariate dummies and interaction terms in Appendix A.  The covariate dummies should follow the standard treatment contrast coding.  So with $J$ strata, use dummy variables $Z_{i2}, \ldots Z_{iJ}$ to encode the strata for unit $i$, giving $J-1$ interaction terms of the form $W_i \cdot (Z_{i2} - \bar{Z}_2), \ldots, W_i \cdot (Z_{iJ} - \bar{Z}_J)$, where $\bar{Z}_j = \frac{1}{N} \sum Z_{ij}$. }

Note that the main results and conclusions regarding conditional randomization tests from \citet{Hennessy2016} do not depend on the proposition.  The proposition was only used in the simulation study to reduce the number of tests to be compared.  Rather than reporting the conditional tests using both $\hat{\tau}_{sd}$ and $\hat{\tau}_{ps}$, we only reported results using $\hat{\tau}_{sd}$.  However, in the specific simulation setting we explored, $\hat{\tau}_{ps}$ is, in fact, a monotonic function of $\hat{\tau}_{sd}$ when conditioning on the observed balance because there are two strata of equal size and the treated and control groups are of equal size.  In this situation, the conditional tests using $\hat{\tau}_{sd}$ and $\hat{\tau}_{ps}$ are equivalent.  We verify this fact below.  For this situation, our test statistics can be expanded as   

\begin{align}
\hat{\tau}_{sd} &= \Big( \frac{N_{T1}}{N_T} \bar{Y}_{T1}^{\text{obs}} + \frac{N_{T2}}{N_T} \bar{Y}_{T2}^{\text{obs}} \Big) - \Big(\frac{N_{C1}}{N_C} \bar{Y}_{C1}^{\text{obs}} + \frac{N_{C2}}{N_C} \bar{Y}_{C2}^{\text{obs}} \Big)  \nonumber \\
  &= \frac{2}{N} ( N_{T1} \bar{Y}_{T1}^{\text{obs}} + N_{T2} \bar{Y}_{T2}^{\text{obs}} - N_{C1} \bar{Y}_{C1}^{\text{obs}} - N_{C2} \bar{Y}_{C2}^{\text{obs}} ) \nonumber
\end{align}

\noindent and 

\begin{align} 
\hat{\tau}_{ps} &= \frac{N_1}{N}(\bar{Y}_{T1}^{\text{obs}} - \bar{Y}_{C1}^{\text{obs}}) + \frac{N_2}{N}(\bar{Y}_{T2}^{\text{obs}} - \bar{Y}_{C2}^{\text{obs}})  \nonumber \\
  &= \frac{1}{2} (\bar{Y}_{T1}^{\text{obs}} + \bar{Y}_{T2}^{\text{obs}} - \bar{Y}_{C1}^{\text{obs}} - \bar{Y}_{C2}^{\text{obs}}) \nonumber .
\end{align} 

\noindent We then show that $\hat{\tau}_{ps}$ is a monotonic function of $\hat{\tau}_{sd}$ by showing that

\begin{align} 
\hat{\tau}_{ps} &= \frac{N}{4 N_{T1} N_{C1}} \Big( \frac{N}{4} (\hat{\tau}_{sd} + 2 \bar{Y}^{\text{obs}}) - N_{T1} \bar{Y}_1^{\text{obs}} - N_{T2} \bar{Y}_2^{\text{obs}} \Big)   \nonumber.
\end{align} 

\noindent The proof is available upon request.

If we were to include the conditional randomization test using $\hat{\tau}_{ps}$ in the simulation study, the results would be the same as those reported for the conditional randomization test using $\hat{\tau}_{sd}$. We leave a formal comparison of conditional randomization tests using $\hat{\tau}_{sd}$ and $\hat{\tau}_{ps}$ for future work.  

\bibliographystyle{plainnat}
\small
\bibliography{references}

\begin{thebibliography}{38}
\providecommand{\natexlab}[1]{#1}
\providecommand{\url}[1]{\texttt{#1}}
\expandafter\ifx\csname urlstyle\endcsname\relax
  \providecommand{\doi}[1]{doi: #1}\else
  \providecommand{\doi}{doi: \begingroup \urlstyle{rm}\Url}\fi

\bibitem[Altman(1985)]{Altman1985}
Douglas~G Altman.
\newblock Comparability of randomised groups.
\newblock \emph{The Statistician}, pages 125--136, 1985.

\bibitem[Angrist and Pischke(2009)]{Angrist2009}
Joshua~D Angrist and Jörn~Steffen Pischke.
\newblock \emph{{Mostly Harmless Econometrics: An Empiricists Companion.}}
\newblock Princeton University Press, 2009.

\bibitem[Birnbaum(1962)]{Birnbaum1962}
Allan Birnbaum.
\newblock On the foundations of statistical inference.
\newblock \emph{Journal of the American Statistical Association}, 57\penalty0
  (298):\penalty0 269--306, 1962.

\bibitem[Bradley(1968)]{Bradley1968}
James~V Bradley.
\newblock Distribution-free statistical tests.
\newblock 1968.

\bibitem[Cox(1958{\natexlab{a}})]{Cox1958}
David~R Cox.
\newblock Some problems connected with statistical inference.
\newblock \emph{Ann. Math. Statist}, 29\penalty0 (2):\penalty0 357--372,
  1958{\natexlab{a}}.

\bibitem[Cox(1958{\natexlab{b}})]{cox1958planning}
David~R Cox.
\newblock \emph{{Planning of experiments.}}
\newblock Wiley, 1958{\natexlab{b}}.

\bibitem[Cox(1982)]{Cox1982}
David~R Cox.
\newblock A remark on randomization in clinical trials.
\newblock \emph{Utilitas Mathematica A}, 21:\penalty0 245--252, 1982.

\bibitem[Ding(2014)]{Ding2014}
Peng Ding.
\newblock A paradox from randomization-based causal inference.
\newblock \emph{http://arxiv.org/abs/1402.0142}, 2014.

\bibitem[Fisher(1956)]{Fisher1956}
R.~A. Fisher.
\newblock \emph{Statistical Methods and Scientific Inference}.
\newblock Oliver \& Boyd, 1956.

\bibitem[Fisher(1935)]{Fisher1935}
Ronald~Aylmer Fisher.
\newblock The design of experiments.
\newblock 1935.

\bibitem[Gail et~al.(1988)Gail, Tan, and Piantadosi]{Gail1988}
MH~Gail, WY~Tan, and S~Piantadosi.
\newblock Tests for no treatment effect in randomized clinical trials.
\newblock \emph{Biometrika}, 75\penalty0 (1):\penalty0 57--64, 1988.

\bibitem[Helland(1995)]{Helland1995}
Inge~S Helland.
\newblock Simple counterexamples against the conditionality principle.
\newblock \emph{The American Statistician}, 49\penalty0 (4):\penalty0 351--356,
  1995.

\bibitem[Hennessy et~al.(2016)Hennessy, Dasgupta, Miratrix, Pattanayak, and
  Sarkar]{Hennessy2016}
Jonathan Hennessy, Tirthankar Dasgupta, Luke Miratrix, Cassandra Pattanayak,
  and Pradipta Sarkar.
\newblock A conditional randomization test to account for covariate imbalance
  in randomized experiments.
\newblock \emph{Journal of Causal Inference}, 4\penalty0 (1):\penalty0 61--80,
  2016.

\bibitem[Huang(1997)]{Huang1997a}
Zhexue Huang.
\newblock A fast clustering algorithm to cluster very large categorical data
  sets in data mining.
\newblock In \emph{DMKD}. Citeseer, 1997.

\bibitem[Iacus et~al.(2012)Iacus, King, and Porro]{Iacus2012}
Stefano~M Iacus, Gary King, and Giuseppe Porro.
\newblock Causal inference without balance checking: Coarsened exact matching.
\newblock \emph{Political Analysis}, 20\penalty0 (1):\penalty0 1--24, 2012.

\bibitem[Kalbfleisch(1975)]{Kalbfleisch1975}
John~D Kalbfleisch.
\newblock Sufficiency and conditionality.
\newblock \emph{Biometrika}, 62\penalty0 (2):\penalty0 251--259, 1975.

\bibitem[Kempthorne(1952)]{Kempthorne1952}
Oscar Kempthorne.
\newblock The design and analysis of experiments.
\newblock 1952.

\bibitem[Kiefer(1977)]{Kiefer1977}
Jack Kiefer.
\newblock Conditional confidence statements and confidence estimators.
\newblock \emph{Journal of the American Statistical Association}, 72\penalty0
  (360a):\penalty0 789--808, 1977.

\bibitem[Kruskal and Wallis(1952)]{Kruskal1952}
William~H Kruskal and W~Allen Wallis.
\newblock Use of ranks in one-criterion variance analysis.
\newblock \emph{Journal of the American statistical Association}, 47\penalty0
  (260):\penalty0 583--621, 1952.

\bibitem[McCane and Albert(2008)]{McCane2008}
Brendan McCane and Michael Albert.
\newblock Distance functions for categorical and mixed variables.
\newblock \emph{Pattern Recognition Letters}, 29\penalty0 (7):\penalty0
  986--993, 2008.

\bibitem[Miratrix et~al.(2013)Miratrix, Sekhon, and Yu]{Miratrix2013}
Luke~W Miratrix, Jasjeet~S Sekhon, and Bin Yu.
\newblock {Adjusting treatment effect estimates by post-stratification in
  randomized experiments}.
\newblock \emph{Journal of the Royal Statistical Society Series B}, 75\penalty0
  (2):\penalty0 369--396, 2013.

\bibitem[Morgan and Rubin(2012)]{Morgan2012}
Kari~Lock Morgan and Donald~B Rubin.
\newblock Rerandomization to improve covariate balance in experiments.
\newblock \emph{The Annals of Statistics}, 40\penalty0 (2):\penalty0
  1263--1282, 2012.

\bibitem[Pattanayak(2011)]{Pattanayak2011}
Cassandra~Wolos Pattanayak.
\newblock \emph{{The Critical Role of Covariate Balance in Causal Inferencewith
  Randomized Experiments and Observational Studies}}.
\newblock PhD thesis, 2011.

\bibitem[Pitman(1938)]{Pitman1938}
Edwin~JG Pitman.
\newblock Significance tests which may be applied to samples from any
  populations: {III.} {The} analysis of variance test.
\newblock \emph{Biometrika}, pages 322--335, 1938.

\bibitem[Raz(1990)]{Raz1990}
Jonathan Raz.
\newblock Testing for no effect when estimating a smooth function by
  nonparametric regression: {A} randomization approach.
\newblock \emph{Journal of the American Statistical Association}, 85\penalty0
  (409):\penalty0 132--138, 1990.

\bibitem[Rosenbaum(1984)]{Rosenbaum1984}
Paul~R Rosenbaum.
\newblock {Conditional permutation tests and the propensity score in
  observational studies}.
\newblock \emph{Journal of the American Statistical Association}, 79\penalty0
  (387):\penalty0 565--574, 1984.

\bibitem[Rosenbaum(2002)]{Rosenbaum2002}
Paul~R Rosenbaum.
\newblock Covariance adjustment in randomized experiments and observational
  studies.
\newblock \emph{Statistical Science}, 17\penalty0 (3):\penalty0 286--304, 2002.

\bibitem[Rubin(1974)]{Rubin:1974wx}
D~B Rubin.
\newblock {Estimating causal effects of treatments in randomized and
  nonrandomized studies}.
\newblock \emph{Journal of Educational Psychology}, 66\penalty0 (5):\penalty0
  688--701, 1974.

\bibitem[Rubin(1980{\natexlab{a}})]{Rubin1980}
Donald~B Rubin.
\newblock Comment.
\newblock \emph{Journal of the American Statistical Association}, 75\penalty0
  (371):\penalty0 591--593, 1980{\natexlab{a}}.

\bibitem[Rubin(1980{\natexlab{b}})]{rubin1980randomization}
Donald~B Rubin.
\newblock {Randomization analysis of experimental data: The Fisher
  randomization test comment}.
\newblock \emph{Journal of the American Statistical Association}, 75\penalty0
  (371):\penalty0 591--593, 1980{\natexlab{b}}.

\bibitem[Rubin(2007)]{Rubin2007}
Donald~B Rubin.
\newblock The design versus the analysis of observational studies for causal
  effects: parallels with the design of randomized trials.
\newblock \emph{Statistics in Medicine}, 26\penalty0 (1):\penalty0 20--36,
  2007.

\bibitem[Senn(1989)]{Senn1989}
SJ~Senn.
\newblock Covariate imbalance and random allocation in clinical trials.
\newblock \emph{Statistics in Medicine}, 8\penalty0 (4):\penalty0 467--475,
  1989.

\bibitem[Splawa-Neyman et~al.(1990)Splawa-Neyman, Dabrowska, and
  Speed]{SplawaNeyman:1990ux}
Jerzy Splawa-Neyman, D~M Dabrowska, and T~P Speed.
\newblock {On the Application of Probability Theory to Agricultural
  Experiments. Essay on Principles. Section 9}.
\newblock \emph{Statistical Science}, 5\penalty0 (4):\penalty0 465--472, 1990.

\bibitem[Stephens et~al.(2013)Stephens, Tchetgen, De~Gruttola,
  et~al.]{Stephens2013}
Alisa~J Stephens, Eric J~Tchetgen Tchetgen, Victor De~Gruttola, et~al.
\newblock Flexible covariate-adjusted exact tests of randomized treatment
  effects with application to a trial of {HIV} education.
\newblock \emph{The Annals of Applied Statistics}, 7\penalty0 (4):\penalty0
  2106--2137, 2013.

\bibitem[Tukey(1993)]{Tukey1993}
John~W Tukey.
\newblock Tightening the clinical trial.
\newblock \emph{Controlled Clinical Trials}, 14\penalty0 (4):\penalty0
  266--285, 1993.

\bibitem[Wilson and Martinez(1997)]{Wilson1997}
D.~Randall Wilson and Tony~R. Martinez.
\newblock Improved heterogeneous distance functions.
\newblock \emph{arXiv preprint cs/9701101}, 1997.

\bibitem[Zhang et~al.(2008)Zhang, Tsiatis, and Davidian]{Zhang2008}
Min Zhang, Anastasios~A Tsiatis, and Marie Davidian.
\newblock Improving efficiency of inferences in randomized clinical trials
  using auxiliary covariates.
\newblock \emph{Biometrics}, 64\penalty0 (3):\penalty0 707--715, 2008.

\bibitem[Zheng and Zelen(2008)]{Zheng2008}
Lu~Zheng and Marvin Zelen.
\newblock Multi-center clinical trials: Randomization and ancillary statistics.
\newblock \emph{The Annals of Applied Statistics}, pages 582--600, 2008.

\end{thebibliography}

\end{document}